# Observation of superconducting pair density modulation within lattice unit cell

Tianheng Wei[1#], Yanzhao Liu[2#], Wei Ren[1#], Zhen Liang[1,5], Ziqiang Wang[3] & Jian Wang[1,4,5*]

[1]*International Center for Quantum Materials, School of Physics, Peking University, Beijing 100871, China*

[2]*Quantum Science Center of Guangdong–Hong Kong–Macao Greater Bay Area (Guangdong), Shenzhen 518045, China*

[3]*Department of Physics, Boston College, Chestnut Hill, MA 02467, USA*

[4]*Collaborative Innovation Center of Quantum Matter, Beijing 100871, China*

[5]*Hefei National Laboratory, Hefei 230088, China*

[#]These authors contributed equally.

*Corresponding to: jianwangphysics@pku.edu.cn (J.W.)

In unconventional high-temperature (high-$T_c$) superconductors, the pair density wave state, an exotic superconducting order showing spatially periodic order parameter modulations with the period of several unit cells and translational symmetry breaking, has attracted broad attention. However, the superconducting pair density modulation within a single unit cell (PDM) has never been carefully investigated before. Here, using scanning tunneling microscopy/spectroscopy, we report the observation of PDM in monolayer high-$T_c$ iron chalcogenide films epitaxially grown on SrTiO$_3$(001). The superconductivity modulations are characterized by the superconducting gap size and the coherence peak sharpness. Further analysis shows that the local maxima and minima in the superconducting gap modulation are centered at the crystallographic locations of the chalcogen atoms, revealing the breaking of the glide-mirror symmetry of the chalcogen atoms in monolayer high-$T_c$ iron chalcogenide films grown on SrTiO$_3$(001). Our findings provide precise microscopic information on superconductivity within the lattice unit cell and may promote the understanding of unconventional high-$T_c$ superconductivity.

*1. Introduction.* Superconductivity in quantum materials, irrespective of whether the Cooper pairing on the Fermi surface is mediated by phonons or electronic fluctuations, is described based on the Bardeen–Cooper–Schrieffer (BCS) theory of Cooper pair condensation on the crystal lattice of the superconductors. In a common uniform superconductor, the superconducting properties are invariant going from one unit cell, which generally contains a number of atoms and atomic orbitals, to another unit cell following the lattice translational symmetry. In recent years, a novel form of superconducting state known as the pair density wave (PDW) order has been theoretically proposed [1–9] and experimentally investigated [10–19]. A PDW order is formed by Cooper pairs with non-zero center-of-mass momentum $\mathbf{Q}$, and breaks the lattice translational symmetry with a spatially modulated superconducting order parameter. The experimental evidence of PDW, albeit coexisting with the uniform superconductivity, has been detected in the high-$T_c$ cuprates [10–13], iron-based superconductors [16,17], and other unconventional superconductors [15,18,19] with the modulation period spanning several times the lattice constant. However, the intra-unit-cell superconducting pair density modulation, referred to as PDM, has not been carefully investigated before. Given the rather local pairing interactions in short coherence length unconventional superconductors [20] and unconventional pairing symmetry [21,22], probing and resolving the superconducting modulations inside a unit cell in such superconductors can provide valuable insights into the nature of superconductivity and inner workings of the pairing mechanism.

Here, by using scanning tunneling microscopy/spectroscopy (STM/S), we report the PDM in monolayer Fe(Te,Se) and FeSe films grown on $SrTiO_3(001)$ (STO) substrates, which are two-dimensional high-$T_c$ iron-based superconductors. The superconductivity modulation is characterized by the superconducting gap size and the coherence peak sharpness modulations.

*2. Results.* The one-unit-cell (1-UC) thick Fe(Te,Se) films were grown by molecular beam epitaxy (MBE) on STO substrates. Figure 1(a) shows the crystal structure of 1-UC Fe(Te,Se) consisting of an Fe layer sandwiched between two Te/Se layers. One primary unit cell (marked by orange in Fig. 1(a)) contains two Fe atoms in the middle layer and two Te/Se atoms in the top and bottom layers. The topmost Te/Se layer exhibits a square lattice structure which is visible in the STM topography. The STM/S measurements were mainly performed at 4.3 K, which is much lower than the superconducting transition temperature (around 60 K) of the monolayer Fe(Te,Se) films on STO [23,24]. Figure 1(b) shows the atomically resolved STM topographic image of the 1-UC FeTe$_{1-x}$Se$_x$ film. The nominal stoichiometry $x \approx 0.7$ is estimated from the thickness of the 2$^{nd}$ layer (around 0.58 nm, Fig. S1) [25]. In Fig. 1(c), the sites of Fe and Te/Se atoms in the Fe(Te,Se) film are marked within a unit cell. To gain more electronic information within the unit cell, we compared the tunneling spectra obtained at the topmost Te/Se sites and the Fe sites. Two typical spectra are plotted in Fig. 1(d). Both spectra are U-shaped with two pairs of coherence peaks, indicating the fully gapped superconductivity with two superconducting gap sizes $\Delta_1$ and $\Delta_2$, which is consistent with previous reports [16,23–25]. Strikingly, the two superconducting gap sizes on the Fe site are larger than those on the topmost Te/Se site, suggesting an intra-unit-cell superconducting gap variation in the 1-UC Fe(Te,Se) film.

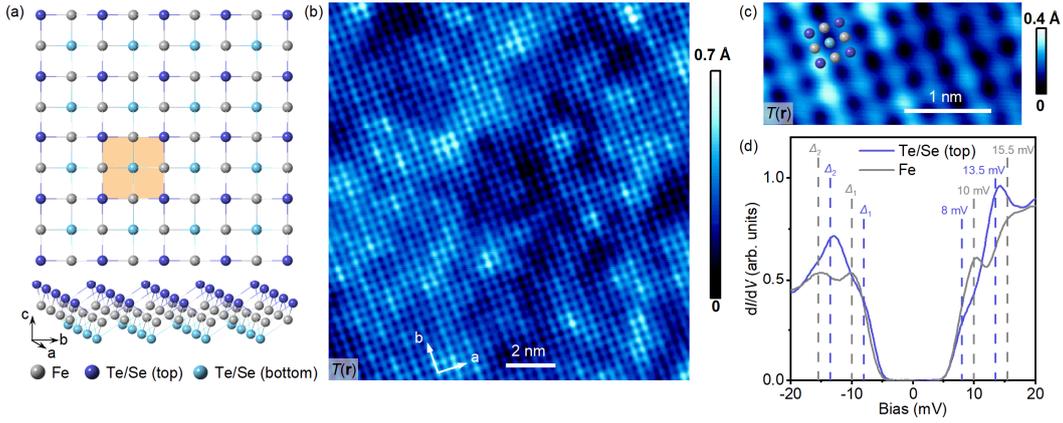

**Fig. 1.** STM characterization of the 1-UC Fe(Te,Se)/STO film (Sample 1). (a) The crystal structure of the 1-UC Fe(Te,Se) film. One primary unit cell is marked by orange, which contains two Fe atoms in the middle layer and two Te/Se atoms in the top and bottom layers. (b) An atomically resolved topographic image of the 1-UC Fe(Te,Se) film (14.5×15.3 nm$^2$, $V_s$ = 40 mV, $I_s$ = 500 pA). (c) A zoom-in of the topographic image of the 1-UC Fe(Te,Se) film. The sites of Fe and Te/Se atoms within one unit cell are marked by colored balls. (d) Typical tunneling spectra measured on the topmost Te/Se site and Fe site ($T$ = 4.3 K, $V_s$ = 40 mV, $I_s$ = 500 pA, $V_{mod}$ = 0.8 mV). Both spectra show fully gapped superconductivity and two pairs of coherence peaks. The superconducting gap sizes $\Delta_1$ and $\Delta_2$ on the Fe site ($\Delta_1 \approx$ 10 mV, $\Delta_2 \approx$ 15.5 mV) are larger than those on the topmost Te/Se site ($\Delta_1 \approx$ 8 mV, $\Delta_2 \approx$ 13.5 mV), suggesting an intra-unit-cell superconducting gap variation. The superconducting gap sizes are determined by half of the distance between the coherence peak positions at positive and negative bias voltages. The slight particle-hole asymmetry may originate from the asymmetric background in the tunneling spectra.

We now focus on the intra-unit-cell superconducting gap variation. High-spatial-resolution tunneling spectra $g(V) \equiv \mathrm{d}I/\mathrm{d}V(V)$ (Fig. 2(b)) are measured along the topmost Te/Se lattice (red arrow in Fig. 2(a)). The tunneling spectra show clear two-gap features $\Delta_1$ and $\Delta_2$ marked by two arrows, respectively. To determine the superconducting gap size $\Delta_1$ and $\Delta_2$ more precisely, Fig. 2(c) shows the color map of the $D(V) \equiv -\mathrm{d}^2g/\mathrm{d}V^2$ calculated from the original measured tunneling spectra shown in Fig. 2(b), in which the bias positions of local maxima in $D(V)$ indicate the superconducting gap sizes [11,16] (see Supplementary Materials for details). The gap positions extracted by this method show consistent modulations with those extracted directly from the coherence peaks in $g(V)$ curves (Fig. S3 and Fig. S8) and the results obtained by the Dynes function fit [26,27] (Fig. S6), which confirms the reliability of this method. As illustrated in Fig. 2(c), both superconducting gap sizes $\Delta_1$ and $\Delta_2$ exhibit spatial modulations along the top Te/Se lattice. The period of the superconducting gap size modulation is equal to the Te/Se lattice constant $a_{Te/Se}$, revealing the intra-unit-cell superconducting gap modulation. Figure 2(d) shows the extracted $\Delta_1$ and $\Delta_2$ values along the distance in Fig. 2(c), which directly show the spatial modulation of superconducting gap size along the top Te/Se lattice (red arrow in Fig. 2(a)). The superconducting gap modulation with a period of $\sqrt{2}a_{Te/Se}$ can also be detected along a linecut in the direction connecting the top and bottom Te/Se atoms (Fig. S15). In addition, the robust intra-unit-cell superconducting gap modulations survive and are discernable along a line cut across a vortex in a high magnetic field (Fig. S11). Moreover, the local $D$ maximum values (purple or blue colors in Fig. 2(c)) reflect the sharpness of superconducting coherence peaks, which can also serve as a local indicator of superconductivity [11]. As shown in Fig. 2(c) and 2(d), the superconducting coherence peak sharpness $D_1$ or $D_2$ for superconducting gap $\Delta_1$ or $\Delta_2$ also exhibits spatial modulation with the period of $a_{Te/Se}$, further confirming the existence of PDM.

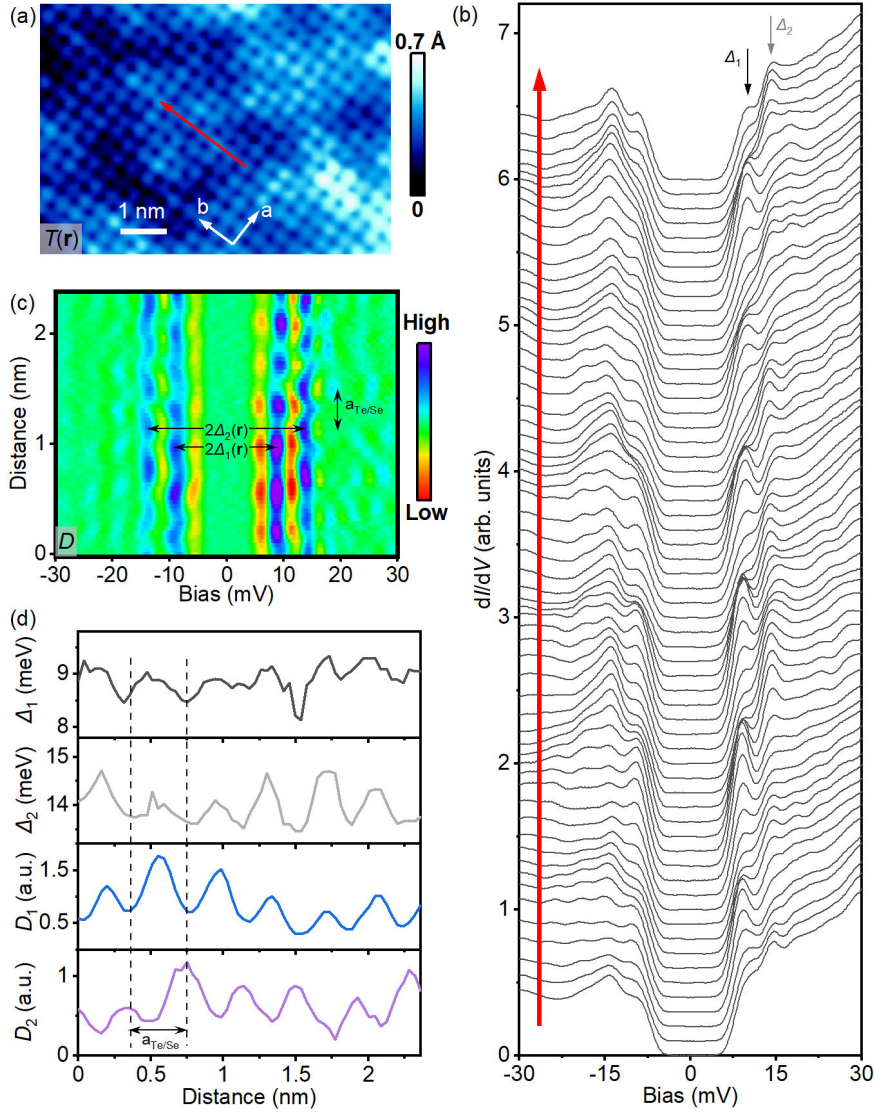

**Fig. 2.** Superconducting gap and coherence peak sharpness modulations with the period of $a_{Te/Se}$ in the 1-UC Fe(Te,Se) film. (a) A topographic image of the 1-UC Fe(Te,Se)/STO film (Sample 1, 7.6×5.5 nm$^2$, $V_s$ = 40 mV, $I_s$ = 500 pA). (b) Tunneling spectra measured along the red arrow in (a) ($T$ = 4.3 K, $V_s$ = 40 mV, $I_s$ = 500 pA, $V_{mod}$ = 0.8 mV). The curves are vertically shifted for clarity. The two arrows indicate the two superconducting gaps $\Delta_1$ and $\Delta_2$. (c) Color map of $D(V) \equiv -d^2g/dV^2$ calculated from (b), which exhibits the spatially modulated superconducting gap sizes ($\Delta_1$ and $\Delta_2$) and coherence peak sharpness ($D_1$ and $D_2$). (d) The extracted superconducting gap sizes ($\Delta_1$ and $\Delta_2$) and coherence peak sharpness ($D_1$ and $D_2$) along the distance in (c). The modulation ratio of the gap sizes $\delta\Delta/\Delta$ is around 7%. All curves show spatial modulation with the period of the Te/Se lattice constant $a_{Te/Se}$. The distances in (c) and (d) are defined relative to the beginning of the arrow in (a).

To gain further insights into the PDM, we carried out spectroscopic imaging STM measurements on a 1-UC Fe(Te,Se) film (Sample 2) prepared with the same recipe as Sample 1. In the region shown in Fig. 3(a), we extracted superconducting gap sizes $\Delta_1$ and $\Delta_2$ from the tunneling spectrum measured at every pixel (the density of pixels is as high as 625 pixels/nm$^2$) and obtained the superconducting gap maps $\Delta_1(\mathbf{r})$ and $\Delta_2(\mathbf{r})$ (Figs. 3(b) and 3(c)). The Fourier transform maps $|\Delta_1(\mathbf{q})|$ and $|\Delta_2(\mathbf{q})|$ (insets of Figs. 3(b) and 3(c)) show distinct Fourier peaks at $\mathbf{Q}_{a,b} = (\pm 1, 0)Q_{Te/Se}$ and $(0, \pm 1)Q_{Te/Se}$, which are equal to the reciprocal lattice vectors, further verifying the intra-unit-cell superconducting gap modulations with the period of $a_{Te/Se}$. After applying the Fourier filtering (Supplementary Materials), the intra-unit-cell modulations of $\Delta_1$ and $\Delta_2$ become clearer (Figs. 3(d) and 3(e)).

In Figs. 3(a)-3(e), the black circles mark the topmost Te/Se sites within a single unit cell, showing the superconducting gap modulations across the crystal lattice. The local minima and maxima of the modulation are centered at crystallographic locations of the topmost and bottommost Te/Se atoms, respectively (Figs. 3(b)-3(e)). The superconducting gap modulations are further confirmed at ultralow temperatures below 100 mK (Figs. S7-S8 and Fig. S20). The different superconducting gap sizes at the topmost and bottommost Te/Se sites reveal the glide-mirror symmetry breaking effect of the Te/Se atoms in 1-UC Fe(Te,Se)/STO, indicating the important role played by the Te/Se atoms in Cooper pairing. Interestingly, the spatial modulations of the small and large superconducting gaps, $\Delta_1(\mathbf{r})$ and $\Delta_2(\mathbf{r})$, exhibit an in-phase relation and reach the minima at topmost Te/Se sites simultaneously. Moreover, the superconducting gap modulations display an antiphase relation with the atomic topography height of the topmost Te/Se lattice plane. The antiphase relation between the superconducting gap and top Te/Se sites is observed in another region in the 1-UC Fe(Te,Se)/STO film (Fig. S18). To quantitatively reveal the phase relationship in the whole field of view, we analyzed the relative phase between the superconducting gap ($\Delta_1$ and $\Delta_2$) modulations and atomic topography by using the 2D lock-in method [12,13,16,28] (see Supplementary Materials for more details). Figure 3(f) shows the distributions of relative phase $\delta\phi_{\mathbf{Q}_{i=a,b}}^{\Delta_{1,2},T}(\mathbf{r})$ between the phase of superconducting gap modulation $\phi_{\mathbf{Q}_i}^{\Delta_{1,2}}(\mathbf{r})$ and the phase of the atomic topography $\phi_{\mathbf{Q}_i}^{T}(\mathbf{r})$ in both $a$ and $b$ directions. Both relative phase distributions (black curve for $\Delta_1$ and red curve for $\Delta_2$) are closer to $\pm\pi$ than to 0, showing the roughly in-phase relation between $\Delta_1$ and $\Delta_2$ modulation and the antiphase relation between the superconducting gap modulations and the atomic topography.

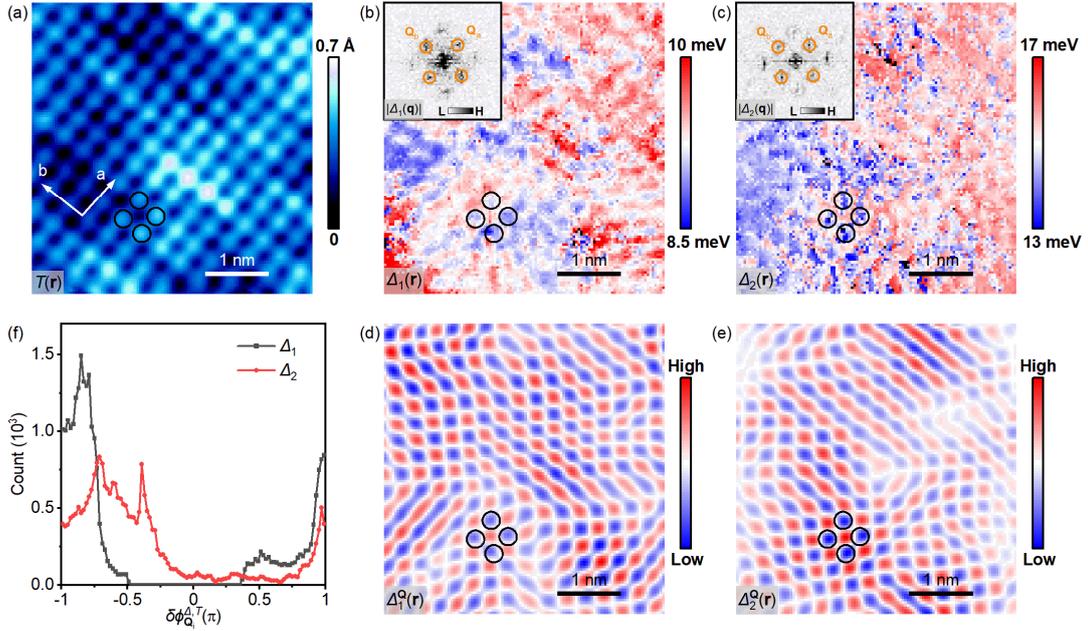

**Fig. 3.** Intra-unit-cell superconducting gap modulation in the 1-UC Fe(Te,Se)/STO film (Sample 2). (a) A topographic image of Sample 2 showing the atomic structures of the topmost Te/Se lattice (4.4×4.6 nm$^2$, $V_s$ = 40 mV, $I_s$ = 500 pA). (b,c) Superconducting gap maps of $\Delta_1$ (b) and $\Delta_2$ (c) measured in the same area as in (a), which show the intra-unit-cell superconducting gap modulations ($T$ = 4.3 K, $V_s$ = 40 mV, $I_s$ = 500 pA, $V_{mod}$ = 0.8 mV, 625 pixels/nm$^2$). The insets of (b) and (c) are the magnitude of the Fourier transform of (b) and (c), respectively. The intra-unit-cell superconducting gap modulation wavevectors ($\mathbf{Q}_a$ and $\mathbf{Q}_b$) are denoted by orange circles, which are at the Bragg points of the topmost Te/Se lattice. (d,e) The Fourier filtered gap maps of $\Delta_1$ (d) and $\Delta_2$ (e). The Fourier filtering process is described in Supplementary Materials. The topmost Te/Se sites within one unit cell are marked by black circles in (a)-(e), clearly revealing the antiphase relation between $\Delta_{1,2}(\mathbf{r})$ and $T(\mathbf{r})$. (f) The distributions of relative phase between superconducting gap size ($\Delta_1$ in black and $\Delta_2$ in red) and atomic topography. The relative phase distributions are closer to $\pm\pi$ than to 0, approximately showing the antiphase relation between the gap size and the atomic topography.

Besides the studies on the superconducting gap size, we also extracted and analyzed the coherence peak sharpness maps $D_{1,2}(\mathbf{r})$ in the same area of Fig. 3. As shown in Fig. 4(b) and 4(c), both $D_1(\mathbf{r})$ and $D_2(\mathbf{r})$ maps exhibit clear modulation features. We mark the topmost Te/Se sites with black circles in Fig. 4(b) and 4(c) and it is clear that $D_1$ is locally minimized at top Te/Se sites where $D_2$ peaks. Comparing the modulations in $\Delta_{1,2}(\mathbf{r})$ and $D_{1,2}(\mathbf{r})$ maps, an in-phase relation is observed between superconducting gap $\Delta_1$ and coherence peak sharpness $D_1$, while $\Delta_2$ and $D_2$ show the antiphase relation. This may be related to the anisotropic gap structures of 1-UC Fe(Te,Se)/STO (Supplementary Materials).

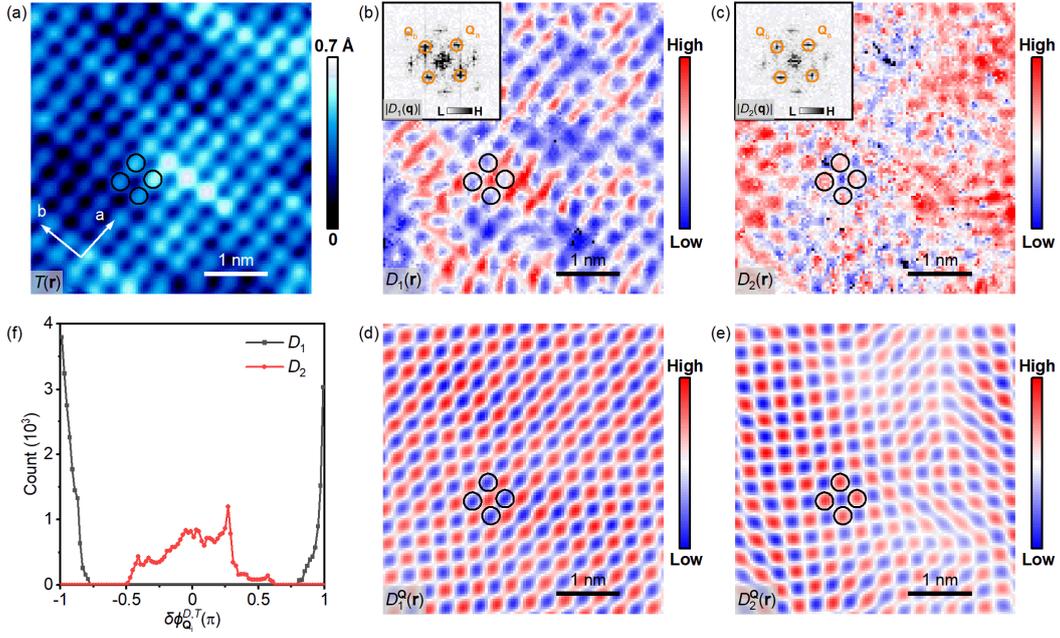

**Fig. 4.** Intra-unit-cell superconducting coherence peak sharpness modulation in the 1-UC Fe(Te,Se) film (Sample 2). (a) Topographic image in Fig. 3(a). (4.4×4.6 nm$^2$, $V_s$ = 40 mV, $I_s$ = 500 pA) (b,c) Superconducting coherence peak sharpness maps of $D_1$ (b) and $D_2$ (c) measured in the same area as in (a), which show the intra-unit-cell $D$ modulations ($T$ = 4.3 K, $V_s$ = 40 mV, $I_s$ = 500 pA, $V_{mod}$ = 0.8 mV). The insets of (b) and (c) are the magnitude of the Fourier transform of (b) and (c), respectively. The intra-unit-cell coherence peak sharpness modulation wavevectors ($\mathbf{Q}_a$ and $\mathbf{Q}_b$) are denoted by orange circles, which are at the Bragg points of the topmost Te/Se lattice. (d,e) The Fourier filtered $D$ map of $D_1$ (d) and $D_2$ (e). The Fourier filter process only keeps the Fourier peaks around $\mathbf{Q}_a$ and $\mathbf{Q}_b$. The topmost Te/Se sites within one unit cell are marked by black circles in (a)-(e). (f) The distributions of the relative phase between superconducting coherence peak sharpness ($D_1$ in black or $D_2$ in red) and atomic topography. The relative phase distributions peak near $\pm\pi$ (0) for $D_1$ ($D_2$), approximately showing antiphase (in-phase) relation between the $D_1$ ($D_2$) and the atomic topography.

To extend our observations of PDM to wider material systems, we grew the monolayer FeSe film on STO substrates (1-UC FeSe/STO) and performed STM measurements. As shown in Fig. 5(a) inset, 1-UC FeSe/STO also shows fully gapped superconductivity with two superconducting gaps $\varDelta_1$ and $\varDelta_2$. By extracting superconducting gaps at every pixel in the region shown in Fig. 5(a), we obtained superconducting gap maps $\varDelta_1(\mathbf{r})$ and $\varDelta_2(\mathbf{r})$ of 1-UC FeSe/STO (Figs. 5(b) and 5(c)). In $\varDelta_1(\mathbf{r})$ and $\varDelta_2(\mathbf{r})$, the intra-unit-cell modulations with the period of $a_{Se}$ are observed, which is further revealed by the Fourier transform maps $|\varDelta_1(\mathbf{q})|$ and $|\varDelta_2(\mathbf{q})|$ (insets of Figs. 5(b) and 5(c)) showing distinct Fourier peaks at $\mathbf{Q}_{a,b}$ = $(\pm1, 0)\mathbf{Q}_{Se}$ and $(0, \pm1)\mathbf{Q}_{Se}$. The Fourier-filtered gap maps (Figs. 5(d) and 5(e)) illustrate the correspondence between the superconducting gap and the crystal lattice. In Figs. 5(b)-5(e), the local minima of the superconducting gap modulation are centered at the top Se sites (black circles in Figs. 5(b)-5(e)), consistent with 1-UC Fe(Te,Se)/STO. This antiphase relation is quantitatively described by the distributions of the relative phase between the phase of superconducting gap modulation and the phase of the atomic topography. As shown in Fig. 5(f), the relative phase distributions for $\varDelta_1$ and $\varDelta_2$ are closer to $\pm\pi$ than to 0, showing the antiphase relation between the superconducting gap modulations and the atomic topography. In addition, similar intra-unit-cell modulation of coherence peak sharpness $D_{1,2}$ is also observed (Fig. S16), further suggesting the PDM in the 1-UC FeSe/STO film. The similar distribution of superconducting gaps within the unit cell (larger at bottom Te/Se sites and smaller at top Te/Se sites) in both 1-UC Fe(Te,Se)/STO and 1-UC FeSe/STO implies that the STO substrate can play an important role in the formation of the PDM in monolayer high-$T_c$ iron chalcogenide films.

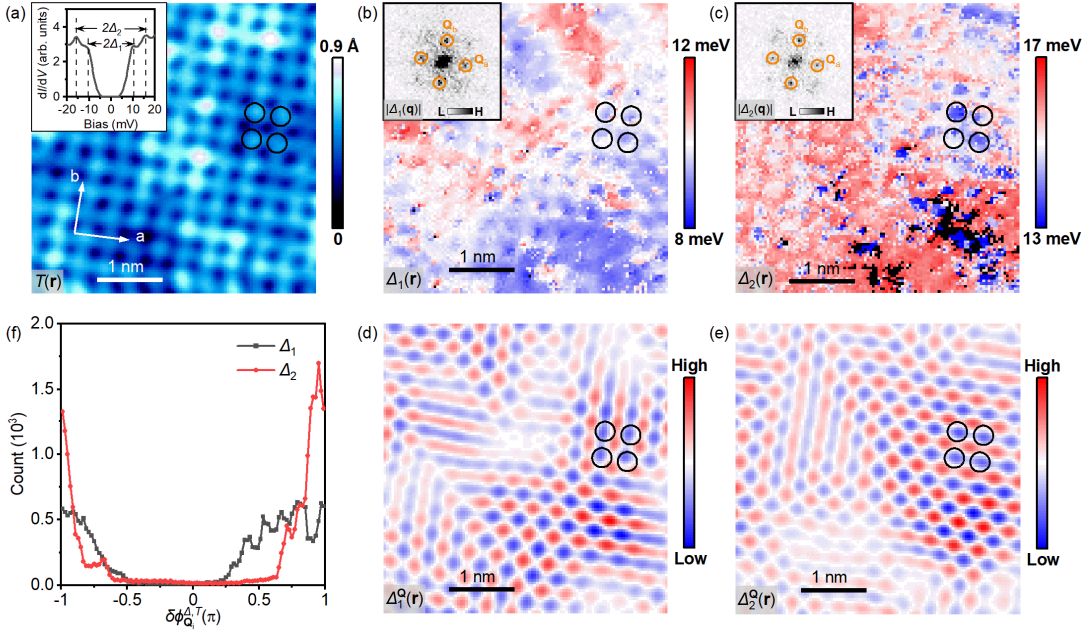

**Fig. 5.** Intra-unit-cell superconducting gap modulation in the 1-UC FeSe/STO film. (a) A topographic image of the 1-UC FeSe/STO showing the atomic structures of the topmost Se lattice ($4.3 \times 4.4$ nm$^2$, $V_s = 40$ mV, $I_s = 500$ pA). The inset shows a typical STS measured on this sample at 4.3 K ($V_s = 40$ mV, $I_s = 2$ nA, $V_{mod} = 0.8$ mV). (b,c) Superconducting gap maps of $\Delta_1$ (b) and $\Delta_2$ (c) measured in the same area as in (a), which show the intra-unit-cell superconducting gap modulations. The insets of (b) and (c) are the magnitude of the Fourier transform of (b) and (c), respectively. The intra-unit-cell superconducting gap modulation wavevectors ($\mathbf{Q}_a$ and $\mathbf{Q}_b$) are denoted by orange circles, which are at the Bragg points of the topmost Se lattice. (d,e) The Fourier filtered gap map of $\Delta_1$ (d) and $\Delta_2$ (e). The topmost Se sites within one unit cell are marked by black circles in (a)-(e), clearly revealing the antiphase relation between $\Delta_{1,2}(\mathbf{r})$ and $T(\mathbf{r})$. (f) The distributions of relative phase between superconducting gap size ($\Delta_1$ in black or $\Delta_2$ in red) and atomic topography. The relative phase distributions are closer to $\pm\pi$ than to 0, approximately showing the antiphase relation between the gap size and the atomic topography. The measurement conditions for the $g(V)$ maps are $T = 4.3$ K, $V_s = 40$ mV, $I_s = 500$ pA, $V_{mod} = 0.8$ mV, 625 pixels/nm$^2$.

*3. Discussion.* Theoretically, in unconventional superconductors, potential scattering caused by impurities could locally modulate the superconductivity [29]. Recent theoretical work reports that pair-breaking scattering interference, which is caused by impurities or Zeeman fields, could also induce superconductivity modulations [30]. However, the superconductivity modulations induced by these scattering effects should have a wavevector of $2k_F$ ($k_F$ is the Fermi wavevector of normal state) [29–31], which is significantly different from the observed intra-unit-cell superconducting gap modulation wavevector at $\mathbf{Q}_{a(b)}$ in 1-UC Fe(Te,Se)/STO and 1-UC FeSe/STO. Thus, the observation cannot be attributed to scattering effects caused by impurities or Zeeman fields. Furthermore, the measured tunneling conductance $\mathrm{d}I/\mathrm{d}V$ can be affected by the tunneling matrix element, which may be influenced by interference effects and complicated tunneling paths [32–34]. While the spatial variations of the tunneling matrix element can change the spectral intensities of the tunneling conductance, they do not usually change the size of the superconducting gap. As shown in Figs. 2-3 and Fig. 5, the superconducting gap size is modulated within a relatively small energy range. The tunneling matrix element can be considered independent of energy within the small energy range, and will not affect the gap position. Therefore, it is unlikely that the observed PDM originates from the tunneling matrix element effects.

Compared with the previously reported PDW states with period spanning multiple unit cells, the superconducting gap size modulations in the 1-UC Fe(Te,Se) and FeSe films have the same period as the unit cell of the crystal lattice containing two Fe atoms and two Te/Se atoms. Therefore, the observed PDM does not break the lattice translational symmetry, which is fundamentally different from the translation symmetry breaking PDW order. Specifically, the superconducting gap size shows significant differences between the topmost and bottommost Te/Se atom sites within one unit cell. For a freestanding monolayer film, the two kinds of Te/Se atom sites should be equivalent under glide-mirror reflection operation. However, in our monolayer films, the glide-mirror symmetry breaking is naturally introduced by the STO substrate, which is believed to be vital for the dramatically enhanced high superconducting transition temperature of the monolayer Fe(Te,Se)/STO film family [35–37]. Therefore, the imbalance between the top and bottom Te/Se atoms, induced by the STO substrate, may lead to different pairing strengths and thus the observed PDM. These findings provide microscopic evidence for the importance of the *p*-orbitals of Te/Se atoms in the superconducting pairing interactions, which has not been sufficiently appreciated before (see Supplementary Materials for the detailed discussion). In addition, in most cases, larger superconducting gaps $\Delta_1$ and $\Delta_2$ are observed at the bottom Te/Se sites in contact with the STO substrate, compared to the smaller values at the top Te/Se sites (Fig. 3 and Fig. 5). This indicates a stronger pairing strength at the bottom Te/Se layer, possibly due to assisted pairing from the phonons in the STO substrate [37].

The PDM could be a universal phenomenon on the crystal lattice with a sublattice structure, where the $\mathbf{Q} = \mathbf{G}$ superconductivity cannot be folded to $\mathbf{Q} = \mathbf{0}$ since $\Delta(\mathbf{r}_i) = \Delta_0 \cos(\mathbf{Q} \cdot \mathbf{r}_i + \varphi)$ is not uniform on different sublattice sites $\mathbf{r}_i$ for $\mathbf{Q} = \mathbf{G}$ superconductivity ($\mathbf{G}$ is the reciprocal lattice vector, $\varphi$ is a constant phase, see Supplementary Materials for more details). This describes the PDM directly observed by STM in real space, where a $\mathbf{Q} = \mathbf{G}$ superconductivity component coexists with the uniform zero-momentum component. For a normal BCS superconductor, the superconducting coherence length is enormously longer than the lattice spacing, and it would be hard to observe PDM on the atomic scale. In contrast, for unconventional superconductors with short coherence lengths, like the 1-UC Fe(Te,Se)/STO (around 1.7 nm, Fig. S11), it is more likely to observe such PDM phenomenon and to reveal the pairing properties at the sub-unit-cell scale. Moreover, it is worth noting that if the pairing order parameter exhibits sign changes within a single unit cell, the PDM can also be anticipated. The explicit determination of the possible sign-change superconductivity at the sub-unit-cell scale requires further phase-sensitive experiments, which could be inspired by our work.

To conclude, we reported the discovery of the PDM in monolayer Fe(Te,Se)/STO and monolayer FeSe/STO films. The modulation peaks and valleys of superconducting gaps are centered at the Te/Se atom sites, which might be attributed to the glide-mirror symmetry breaking induced by the STO substrate in such high-$T_c$ superconductors. Our findings provide the precise microscopic visualization of local superconductivity within the lattice unit cell and offer new insights into the unconventional superconductors containing multiple atoms per unit cell.

*Acknowledgement.* We thank Patrick A. Lee for the discussions. This work was supported by the National Natural Science Foundation of China (Grant No. 12488201, 12404215), the Innovation Program for Quantum Science and Technology (2021ZD0302403), Guangdong Provincial Quantum Science Strategic Initiative (GDZX2401001, GDZX2401009). Z.W. is supported by the U.S. Department of Energy, Basic Energy Sciences Grant No. DE-FG02-99ER45747.

# References

[1] A. Himeda, T. Kato, and M. Ogata, Stripe States with Spatially Oscillating $d$-Wave Superconductivity in the Two-Dimensional $t$-$t'$-$J$ Model, Phys. Rev. Lett. **88**, 117001 (2002).

[2] H.-D. Chen, O. Vafek, A. Yazdani, and S.-C. Zhang, Pair Density Wave in the Pseudogap State of High Temperature Superconductors, Phys. Rev. Lett. **93**, 187002 (2004).

[3] E. Berg, E. Fradkin, E.-A. Kim, S. A. Kivelson, V. Oganesyan, J. M. Tranquada, and S. C. Zhang, Dynamical Layer Decoupling in a Stripe-Ordered High-$T_c$ Superconductor, Phys. Rev. Lett. **99**, 127003 (2007).

[4] D. F. Agterberg and H. Tsunetsugu, Dislocations and vortices in pair-density-wave superconductors, Nat. Phys. **4**, 639 (2008).

[5] E. Berg, E. Fradkin, and S. A. Kivelson, Theory of the striped superconductor, Phys. Rev. B **79**, 064515 (2009).

[6] E. Berg, E. Fradkin, and S. A. Kivelson, Charge-4e superconductivity from pair-density-wave order in certain high-temperature superconductors, Nat. Phys. **5**, 830 (2009).

[7] P. A. Lee, Amperean Pairing and the Pseudogap Phase of Cuprate Superconductors, Phys. Rev. X **4**, 031017 (2014).

[8] D. F. Agterberg, J. C. S. Davis, S. D. Edkins, E. Fradkin, D. J. Van Harlingen, S. A. Kivelson, P. A. Lee, L. Radzihovsky, J. M. Tranquada, and Y. Wang, The Physics of Pair-Density Waves: Cuprate Superconductors and Beyond, Annu. Rev. Condens. Matter Phys. **11**, 231 (2020).

[9] E. Fradkin, S. A. Kivelson, and J. M. Tranquada, Colloquium: Theory of intertwined orders in high temperature superconductors, Rev. Mod. Phys. **87**, 457 (2015).

[10] M. H. Hamidian et al., Detection of a Cooper-pair density wave in $Bi_2Sr_2CaCu_2O_{8+x}$, Nature **532**, 343 (2016).

[11] W. Ruan, X. Li, C. Hu, Z. Hao, H. Li, P. Cai, X. Zhou, D.-H. Lee, and Y. Wang, Visualization of the periodic modulation of Cooper pairing in a cuprate superconductor, Nat. Phys. **14**, 1178 (2018).

[12] S. D. Edkins et al., Magnetic field–induced pair density wave state in the cuprate vortex halo, Science **364**, 976 (2019).

[13] Z. Du, H. Li, S. H. Joo, E. P. Donoway, J. Lee, J. C. S. Davis, G. Gu, P. D. Johnson, and K. Fujita, Imaging the energy gap modulations of the cuprate pair-density-wave state, Nature **580**, 65 (2020).

[14] X. Liu, Y. X. Chong, R. Sharma, and J. C. S. Davis, Discovery of a Cooper-pair density wave state in a transition-metal dichalcogenide, Science **372**, 1447 (2021).

[15] H. Chen et al., Roton pair density wave in a strong-coupling kagome superconductor, Nature **599**, 222 (2021).

[16] Y. Liu, T. Wei, G. He, Y. Zhang, Z. Wang, and J. Wang, Pair density wave state in a monolayer high-$T_c$ iron-based superconductor, Nature **618**, 934 (2023).

[17] H. Zhao, R. Blackwell, M. Thinel, T. Handa, S. Ishida, X. Zhu, A. Iyo, H. Eisaki, A. N. Pasupathy, and K. Fujita, Smectic pair-density-wave order in $EuRbFe_4As_4$, Nature **618**, 940 (2023).

[18] A. Aishwarya et al., Magnetic-field-sensitive charge density waves in the superconductor $UTe_2$, Nature **618**, 928 (2023).

[19] Q. Gu et al., Detection of a pair density wave state in $UTe_2$, Nature **618**, 921 (2023).

[20] G. Blatter, M. V. Feigel'man, V. B. Geshkenbein, A. I. Larkin, and V. M. Vinokur, Vortices in high-temperature superconductors, Rev. Mod. Phys. **66**, 1125 (1994).

[21] C. C. Tsuei and J. R. Kirtley, Pairing symmetry in cuprate superconductors, Rev. Mod. Phys. **72**, 969 (2000).

[22] R. M. Fernandes, A. I. Coldea, H. Ding, I. R. Fisher, P. J. Hirschfeld, and G. Kotliar, Iron pnictides and chalcogenides: a new paradigm for superconductivity, Nature **601**, 35 (2022).

[23] C. Chen, C. Liu, Y. Liu, and J. Wang, Bosonic Mode and Impurity-Scattering in Monolayer Fe(Te,Se) High-Temperature Superconductors, Nano Lett. **20**, 2056 (2020).

[24] C. Chen, K. Jiang, Y. Zhang, C. Liu, Y. Liu, Z. Wang, and J. Wang, Atomic line defects and zero-energy end states in monolayer Fe(Te,Se) high-temperature superconductors, Nat. Phys. **16**, 536 (2020).

[25] F. Li et al., Interface-enhanced high-temperature superconductivity in single-unit-cell $FeTe_{1-x}Se_x$ films on $SrTiO_3$, Phys. Rev. B **91**, 220503 (2015).

[26] R. C. Dynes, V. Narayanamurti, and J. P. Garno, Direct Measurement of Quasiparticle-Lifetime Broadening in a Strong-Coupled Superconductor, Phys. Rev. Lett. **41**, 1509 (1978).

[27] Z. Du, X. Yang, H. Lin, D. Fang, G. Du, J. Xing, H. Yang, X. Zhu, and H.-H. Wen, Scrutinizing the double superconducting gaps and strong coupling pairing in $(Li_{1-x}Fe_x)OHFeSe$, Nat. Commun. **7**, 10565 (2016).

[28] K. Fujita et al., Direct phase-sensitive identification of a *d*-form factor density wave in underdoped cuprates, Proc. Natl. Acad. Sci. **111**, E3026 (2014).

[29] A. V. Balatsky, I. Vekhter, and J.-X. Zhu, Impurity-induced states in conventional and unconventional superconductors, Rev. Mod. Phys. **78**, 373 (2006).

[30] Z.-Q. Gao, Y.-P. Lin, and D.-H. Lee, Pair-breaking scattering interference as a mechanism for superconducting gap modulation, Phys. Rev. B **110**, 224509 (2024).

[31] X. Shi, Z.-Q. Han, P. Richard, X.-X. Wu, X.-L. Peng, T. Qian, S.-C. Wang, J.-P. Hu, Y.-J. Sun, and H. Ding, $FeTe_{1-x}Se_x$ monolayer films: towards the realization of high-temperature connate topological superconductivity, Sci. Bull. **62**, 503 (2017).

[32] I. Martin, A. V. Balatsky, and J. Zaanen, Impurity States and Interlayer Tunneling in High Temperature Superconductors, Phys. Rev. Lett. **88**, 097003 (2002).

[33] Y. Chen, T. M. Rice, and F. C. Zhang, Rotational Symmetry Breaking in the Ground State of Sodium-Doped Cuprate Superconductors, Phys. Rev. Lett. **97**, 237004 (2006).

[34] J. Nieminen, H. Lin, R. S. Markiewicz, and A. Bansil, Origin of the Electron-Hole Asymmetry in the Scanning Tunneling Spectrum of the High-Temperature $Bi_2Sr_2CaCu_2O_{8+\delta}$ Superconductor, Phys. Rev. Lett. **102**, 037001 (2009).

[35] Q.-Y. Wang et al., Interface-Induced High-Temperature Superconductivity in Single Unit-Cell FeSe Films on $SrTiO_3$, Chin. Phys. Lett. **29**, 037402 (2012).

[36] W.-H. Zhang et al., Direct Observation of High-Temperature Superconductivity in One-Unit-Cell FeSe Films, Chin. Phys. Lett. **31**, 017401 (2014).

[37] J. J. Lee et al., Interfacial mode coupling as the origin of the enhancement of $T_c$ in FeSe films on $SrTiO_3$, Nature **515**, 245 (2014).

**Supplementary Materials:**

# Observation of superconducting pair density modulation within lattice unit cell

Tianheng Wei[1#], Yanzhao Liu[2#], Wei Ren[1#], Zhen Liang[1,5], Ziqiang Wang[3] & Jian Wang[1,4,5*]

[1]*International Center for Quantum Materials, School of Physics, Peking University, Beijing 100871, China*
[2]*Quantum Science Center of Guangdong–Hong Kong–Macao Greater Bay Area (Guangdong), Shenzhen 518045, China*
[3]*Department of Physics, Boston College, Chestnut Hill, MA 02467, USA*
[4]*Collaborative Innovation Center of Quantum Matter, Beijing 100871, China*
[5]*Hefei National Laboratory, Hefei 230088, China*

[#]These authors contributed equally.
*Corresponding to: jianwangphysics@pku.edu.cn (J.W.)

**Sample growth and STM measurement**

Our experiments were performed on a commercial USM-1600 ultra-high vacuum MBE-STM combined system. The Fe(Te,Se) films were grown according to the following recipe: The Nb-doped SrTiO$_3$(001) (wt 0.7%) substrates were annealed at 1000℃ in the MBE chamber for 1 hour and then cooled down to 330℃ for film growth. High-purity Fe(99.995%, 1030℃), Te(99.9999%, 245℃), and Se(99.999%, 130℃) were co-deposited on substrates from Knudsen cells for 9 minutes. Then the as-grown films were annealed at 400℃ for 17 hours. For the growth of FeSe/STO, the Nb-doped SrTiO$_3$(001) (wt 0.7%) substrate was annealed at 1000℃ in the MBE chamber for 1 hour and then cooled down to 400℃ for film growth. High purity Fe(99.995%, 1030℃) and Se(99.999%, 130℃) were co-deposited on the substrate from Knudsen cells for 9 minutes. Then the as-grown films were annealed at 450℃ for 13 hours. The thickness of the 1-UC FeTe$_{1-x}$Se$_x$ film is around 0.58 nm (Fig. S1), corresponding to the composition x ≈ 0.7 [1]. Because the monolayer Fe(Te,Se)/STO has a relatively large surface corrugation, which is probably due to the roughness of the substrate, the brighter atoms cannot simply be assigned to Te and the darker to Se [2]. Thus, the Te/Se ratio cannot be determined by counting Te/Se directly from the STM topography [3].

After annealing, the films were transported to the *in situ* low-temperature STM kept at 4.3 K. The STM/S measurements were performed with polycrystalline PtIr tips. The STS spectra were acquired by standard lock-in technique at 983 Hz.

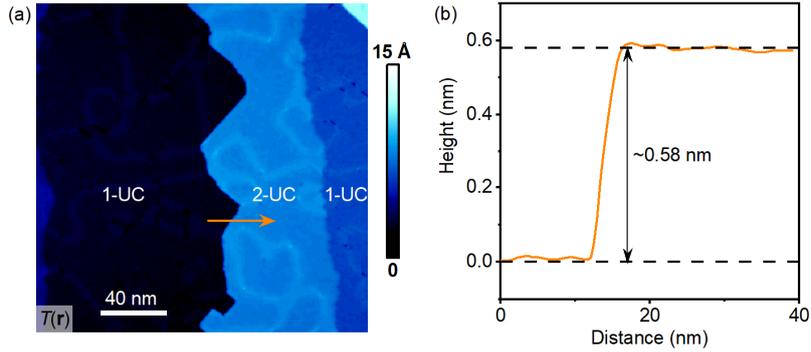

**Fig. S1.** More information about the 1-UC Fe(Te,Se)/STO (Sample 2). (a) A large-scale topographic image of the Fe(Te,Se)/STO film with terraces of STO substrate and partially grown 2$^{nd}$-UC Fe(Te,Se) (200×200 nm$^2$, $V_s$ = 1 V, $I_s$ = 100 pA). The right 1-UC area is grown on a higher terrace and is around 0.4 nm (the height of one SrTiO$_3$ unit cell) higher than the left 1-UC area. (b) The altitude line profile taken along the orange arrow in (a). The thickness of 2$^{nd}$-UC FeTe$_{1-x}$Se$_x$ is around 0.58 nm, corresponding to x ≈ 0.7.

**Extraction of superconducting gap size and coherence peak sharpness**

The $g \equiv \mathrm{d}I/\mathrm{d}V$ value measured by STM is proportional to $|M(\mathbf{r})|^2 N(\mathbf{r}, V)$, where $M(\mathbf{r})$ is the tunneling matrix element and $N(\mathbf{r}, V)$ is the local density of states. For low bias voltages used in tunneling measurements, the $M(\mathbf{r})$ is approximately constant at different energies [4]. Therefore, the superconducting gap energy $\Delta(\mathbf{r})$ extracted from $g(\mathbf{r}, V)$ is irrelevant to the $M(\mathbf{r})$ as $\Delta$ only depends on the shape of the $g(\mathbf{r}, V)$ curve.

As shown in Fig. S2, the coherence peak structures of superconductivity could degrade into shoulder structures. To clearly reveal the peak and shoulder structures with the same method, we took the negative second derivative of the $g(V)$ spectra, namely the $D$ spectra ($D \equiv -\mathrm{d}^2 g/\mathrm{d}V^2$), to characterize the coherence peak structure. The peaks in $D$ spectra correspond to the superconducting gap shown as peak or shoulder structures in $g(V)$ spectra. The larger peak value of $D$ represents sharper coherence peak. Thus, we extracted superconducting gap size and coherence peak sharpness as follows:

a) Calculate the $D$ spectra from the $g(V)$ spectra;

b) Find the bias $V_1^0$ and $V_2^0$ that have local maxima of $D$ near the two superconducting gap energies $\Delta_1$ and $\Delta_2$ extracted from the STS, respectively.

c) For each bias $V_i^0$ ($i = 1,2$), use three data points of $D(V)$ in the neighborhood of $V_i^0$, namely $\{V_i^{-1}, V_i^0, V_i^{+1}\}$, to fit a quadratic function and take the apex position $|eV_i|$ as the gap value $\Delta_i$, and the apex value as the coherence peak sharpness $D_i$.

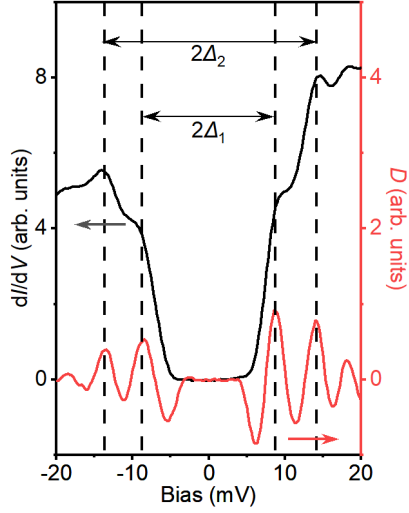

**Fig. S2.** Extraction of superconducting gap. The black curve shows a typical $g(V)$ spectrum with two pairs of superconducting coherence peaks ($T = 4.3$ K, $V_s = 40$ mV, $I_s = 500$ pA, $V_{mod} = 0.8$ mV). The red curve shows the $D \equiv -\mathrm{d}^2g/\mathrm{d}V^2$ spectrum, which shows clearer peak features at the superconducting gap energy marked by black dashed lines.

The reliability of the gap extraction progress is further confirmed by comparing $g(V)$ curves and $D$ curves obtained along a line profile (Fig. S3). The superconducting gap positions show same modulations in both original $g(V)$ curves and $D$ curves.

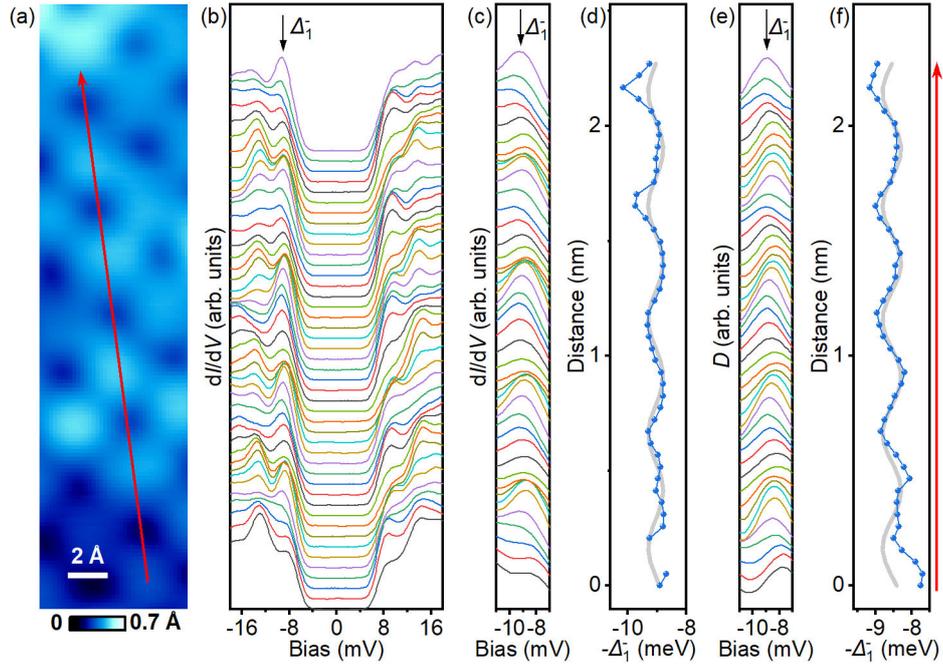

**Fig. S3.** Superconducting gap modulation in $g(V)$ curves and $D$ curves (Sample 3). (a) A topography of the 1-UC Fe(Te,Se)/STO film ($V_s$ = 40 mV, $I_s$ = 500 pA). (b) $g(V)$ curves measured along the red arrow in (a). (c) $g(V)$ curves near the coherence peak around -9 meV. (d) Extracted gap $-\Delta_1^-$ from (c). Some points are missing because there is no peak in corresponding $g(V)$ curves. (e) $D$ curves calculated from $g(V)$ curves in (c). (f) Extracted gap $-\Delta_1^-$ from (e). The blue curves in (d) and (f) show the data and the gray curves show the cosine function with period of $\sqrt{2}a_{\mathrm{Te/Se}}$. The curves in (b), (c), and (e) are vertically shifted for clarity.

Note that the coherence peak sharpness value we used is $D_i$ = -d$^2g$/d$V^2$($\mathbf{r}$, $\Delta_i$), where the bias voltage $V$ is determined by the superconducting gap $\Delta_i$ at position $\mathbf{r}$. Therefore, the $D_i$ depends on the local curvature of $g(V)$ at gap energy, which primarily reflects superconductivity information. As shown in Fig. S4, the $D_i$ = -d$^2g$/d$V^2$($\mathbf{r}$, $\Delta_i$) shows very strong modulation with the lattice periodicity, while the $D$($\mathbf{r}$, $V$) does not show modulation with the lattice for $V$ larger than the superconducting gap energy ($V$ = 30 mV), further supporting the modulation of coherence sharpness originates from the superconductivity modulation.

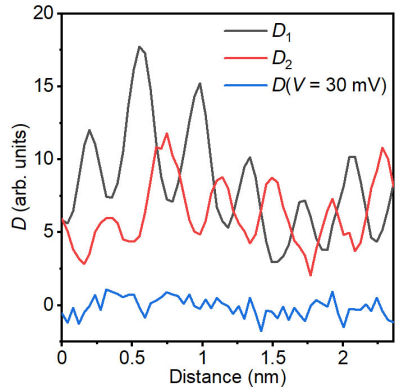

**Fig. S4.** Comparison between $D$ modulations at gap energy and bias larger than gap energy. The $D$ values are extracted from Fig. 2(c). $D_{i=1,2}$ = -d$^2$g/d$V^2$($\mathbf{r}$, $\Delta_i$) show strong lattice periodic modulation, while the $D(\mathbf{r}$, $V$ = 30 mV) does not show the modulation with lattice, suggesting the $D_{i=1,2}$ modulations reflect the modulation of superconductivity.

For both positive (+) and negative (-) bias we can extract $\Delta_{i=1,2}^{+(-)}$ and $D_{i=1,2}^{+(-)}$. The $\Delta$ and $D$ shown in Figs. 2-5 are $\Delta_{i=1,2} = (\Delta_{i=1,2}^+ + \Delta_{i=1,2}^-)/2$ and $D_{i=1,2} = (D_{i=1,2}^+ + D_{i=1,2}^-)/2$. For rare cases in which the gap extraction processes failed, the corresponding pixels were labeled as "bad pixels" and plotted in black in the $\Delta(\mathbf{r})$ and $D(\mathbf{r})$ maps. Figure S2 shows a typical example of the gap extraction results. Fig. S5 and Fig. S17 show the extracted $\Delta_1(\mathbf{r})$, $D_1(\mathbf{r})$ maps at positive ($\Delta_1^+(\mathbf{r})$, $D_1^+(\mathbf{r})$) and negative ($\Delta_1^-(\mathbf{r})$, $D_1^-(\mathbf{r})$) biases in the region shown in Fig. 3. The in-phase relation between $\Delta_1^+(\mathbf{r})$ ($D_1^+(\mathbf{r})$) and $\Delta_1^-(\mathbf{r})$ ($D_1^-(\mathbf{r})$) verifies that the extracted $\Delta_1(\mathbf{r})$ ($D_1(\mathbf{r})$) are symmetric between positive and negative biases.

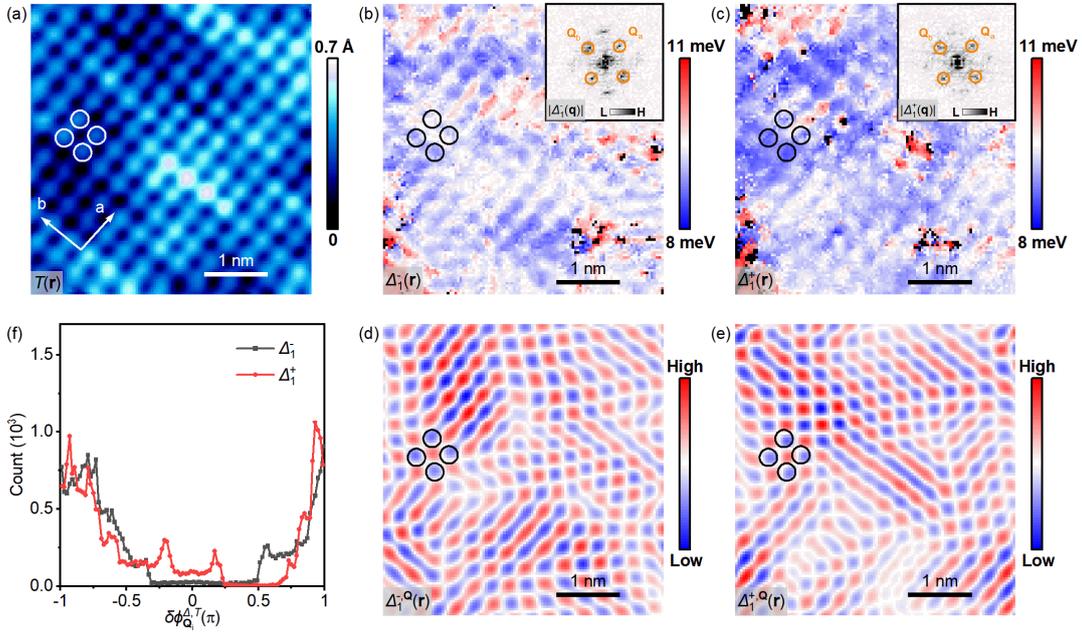

**Fig. S5.** In-phase superconducting gap modulations at negative and positive biases. (a) The topographic image shown in Fig. 3(a) (4.4×4.6 nm$^2$, $V_s$ = 40 mV, $I_s$ = 500 pA). (b,c) Superconducting gap maps of $\Delta_1$ extracted at negative (b) and positive (c) biases in the same area as in (a), which show the intra-unit-cell superconducting gap modulations ($T$ = 4.3 K, $V_s$ = 40 mV, $I_s$ = 500 pA, $V_{mod}$ = 0.8 mV). The insets of (b) and (c) are the magnitude of the Fourier transform of (b) and (c), respectively. (d,e) The Fourier filtered gap map of $\Delta_1^-$ (d) and $\Delta_1^+$ (e). The topmost Te/Se sites within one unit cell are marked by circles in (a)-(e), clearly revealing the antiphase relation between $\Delta_1^{-,+}(\mathbf{r})$ and $T(\mathbf{r})$. (f) The distributions of the relative phase between superconducting gap size ($\Delta_1^-$ in black or $\Delta_1^+$ in red) and atomic topography. The relative phase distributions are closer to ±π than to 0, approximately showing that both $\Delta_1^-$ and $\Delta_1^+$ show antiphase relation with topography. The modulations of $\Delta_1^-$ and $\Delta_1^+$ are in-phase with each other, showing particle-hole symmetric modulations, which excludes the possibility that the gap modulations originate from the asymmetric background.

**Superconducting gap modulation extracted from dynes function fit**

To exclude possible influence from the different coherence peak shapes on the extracted superconducting gap size, we fit the spectra shown in Fig. 2(b) based on the Dynes model [5,6] and extracted the superconducting gap sizes from the fitting parameters. The original d$I$/d$V$ curves are shown in Fig. S6(a), which have a particle-hole asymmetric background originating from the normal density of states. We normalized the original d$I$/d$V$ curves by dividing the cubic fitting to the spectra for $|V| \geq 30$ mV, as shown in Fig. S6(b). Then we fit the normalized spectra with the two-band Dynes model and the fitting curves are shown in Fig. S6(c). Two superconducting gap sizes are extracted from the fitting parameters and show consistent modulations with the gap sizes extracted from the $D$ curves (Figs. S6(d,e)), which confirms the superconducting gap modulations.

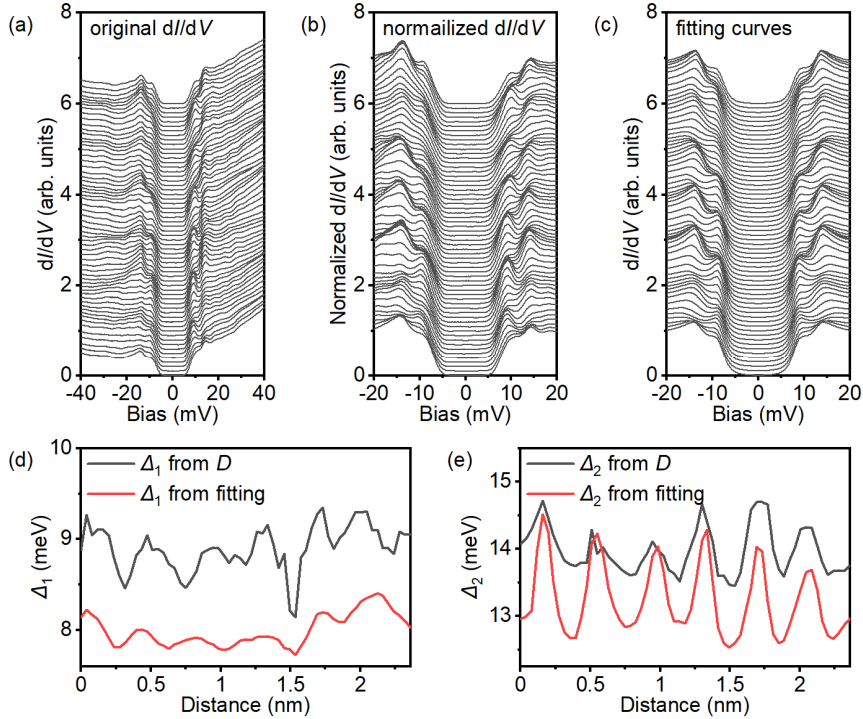

**Fig. S6.** Superconducting gap modulations extracted from fitting parameters. (a) Original tunneling spectra shown in Fig. 2(b). (b) Normalized spectra calculated by dividing the cubic fitting to the original spectra for $|V| \geq 30$ mV. (c) Fitting curves to the normalized spectra based on two-band Dynes model. The curves in (a)-(c) are vertically shifted for clarity. (d,e) Superconducting gap modulations of $\Delta_1$ (d) and $\Delta_2$ (e) extracted from fitting parameters (red curves) in comparison with gap modulations extracted from $D$ (black curves).

**Superconducting gap modulation at ultralow temperatures**

To exclude possible influence from the relatively blunt coherence peak on the extracted superconducting gap size, we performed experiments at ultralow temperatures. We measured the $g(V)$ curves with smaller modulation voltage in the lock-in amplifier and smaller tip-sample distance to achieve higher energy resolution. We obtained sharper coherence peak structure in most areas, which makes it easier to extract the superconducting gap value directly from the $g(V)$ curves without taking further derivatives.

As shown in Fig. S7, we measured $g(V)$ curves along the Te/Se(top)-Te/Se(bottom) direction, which show the clear coherence peak structures. We extracted the superconducting gap energy $\Delta_1$ directly from the $g(V)$ curves as shown in Fig. S7(d). The superconducting gap size $\Delta_1$ shows clear intra-unit-cell modulation.

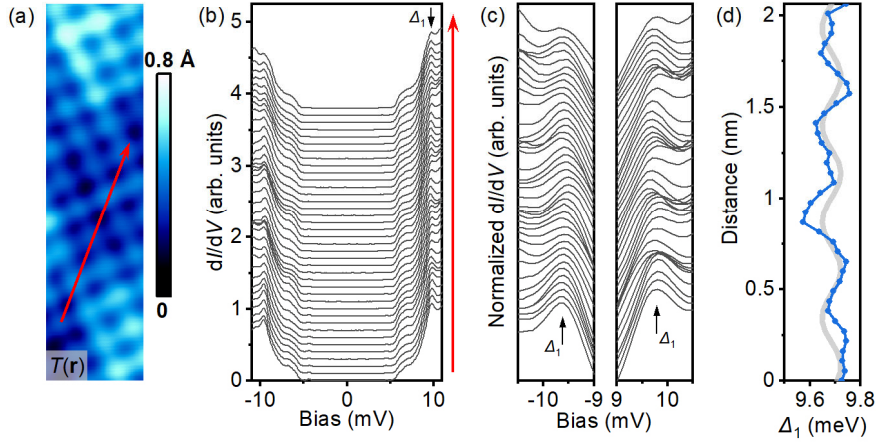

**Fig. S7.** Superconducting gap modulation along Se/Te(top)-Se/Te(bottom) direction in the 1-UC Fe(Te,Se)/STO film at 60 mK (Sample 3). (a) A topography of 1-UC Fe(Te,Se)/STO film ($V_s = 30$ mV, $I_s = 8$ nA). (b) $g(V)$ curves measured along the red arrow in (a) ($T = 60$ mK, $V_s = 30$ mV, $I_s = 8$ nA, $V_{mod} = 0.2$ mV). (c) Normalized $g(V)$ curves near the coherence peak around ±9.5 mV. The $g(V)$ curves shown in the left (right) panel are normalized by dividing $g$ values at 9.5 mV (-9.5 mV). (d) Extracted $\Delta_1$ from the coherence peaks in $g(V)$ curves (blue line). The gray curve is the extracted gap size modulation by applying the Fourier filter. The curves in (b) and (c) are vertically shifted for clarity.

As shown in Fig. S8, we measured the $g(V)$ map in a 3.4×3.3 nm$^2$ field of view (FOV) and extracted the superconducting gap size $\Delta_1$ directly from the original $g(V)$ curves (Fig. S8(b)). For comparison, we also extracted the superconducting gap size $\Delta_1$ using the second derivative $D$ curves as shown in Fig. S8(c). The $\Delta_1(\mathbf{r})$ maps extracted from both methods show almost identical intra-unit-cell modulations and antiphase relations with the atomic topography (Fig. S8(d)), which are consistent with our 4.3 K results. These results exclude the possible influence of the shape of the coherence peak and unambiguously demonstrate the intra-unit-cell superconducting gap modulations.

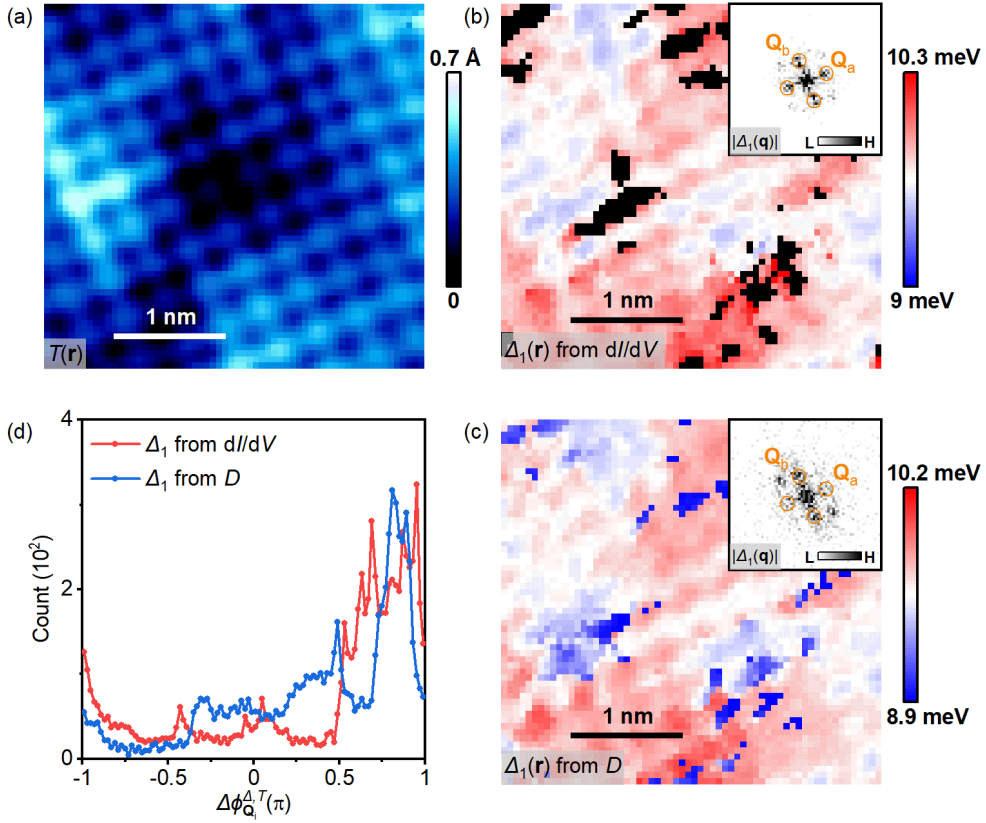

**Fig. S8.** Intra-unit-cell superconducting gap modulation in the 1-UC Fe(Te,Se)/STO film at 60 mK (Sample 3). (a) A topographic image showing the atomic structures of the topmost Te/Se lattice (3.3×3.4 nm$^2$, $V_s = 30$ mV, $I_s = 8$ nA). (b,c) Superconducting gap maps of $\Delta_1$ extracted from $g(V)$ curves (b) and $D$ curves (c) in the same area as in (a), which show the intra-cell superconducting gap modulations ($T = 60$ mK, $V_s = 30$ mV, $I_s = 8$ nA, $V_{mod} = 0.2$ mV, 333 pixels/nm$^2$). The superconducting gaps are extracted from the coherence peak structures at positive bias. Black pixels represent no well-defined peak structures at those positions. The insets of (b) and (c) are the magnitude of the Fourier transform of (b) and (c), respectively. The intra-unit-cell superconducting gap modulation wavevectors ($\mathbf{Q}_a$ and $\mathbf{Q}_b$) are denoted by orange circles, which are at the Bragg points of the topmost Te/Se lattice. (d) The distributions of the relative phase between superconducting gap size $\Delta_1$ (extracted by two methods) and atomic topography. The relative phase distributions are closer to $\pm\pi$ than to 0, approximately showing the antiphase relation between the gap size and the atomic topography.

**Relative phase analysis based on 2D lock-in method**

To quantitatively determine which atomic sites are the modulation extremes centered at, we need to compare the phase information of topography and modulation maps. However, a typical Fourier map only shows the intensity of the Fourier transform and the detailed phase information is absent. To extract the detailed phase information of modulations with wavevector $\mathbf{Q}$ in real space map $C(\mathbf{r})$, we used the 2D lock-in method to calculate the spatial complex amplitudes of the modulations as follows:

$$A_{\mathbf{Q}}^{C}(\mathbf{r}) = F^{-1}\left[ F\left[C(\mathbf{r})e^{i\mathbf{Q}\cdot\mathbf{r}}\right] \cdot \frac{1}{2\pi\sigma_q^2} e^{-\frac{\mathbf{q}^2}{2\sigma_q^2}} \right]$$

where $F$ and $F^{-1}$ are the Fourier transform and the inverse Fourier transform, and $\mathbf{q}$ is the coordinate in Fourier space and $\sigma_q$ is the cut-off length in Fourier space. In practice, we chose $\sigma_q = 1/\lambda$ and $\lambda$ is the period of the modulation ($\lambda = 2\pi/Q$). From the spatial complex amplitude, we can extract its phase as a phase map:

$$\phi_{\mathbf{Q}}^C(\mathbf{r}) = \mathrm{Arg}\left(A_{\mathbf{Q}}^C(\mathbf{r})\right)$$

We obtained phase maps at $\mathbf{Q}_{a,b}$ for atomic topography $T(\mathbf{r})$, superconducting gap map $\Delta_{1,2}(\mathbf{r})$ and superconducting coherence peak sharpness map $D_{1,2}(\mathbf{r})$, respectively. The relative phase $\delta\phi_{\mathbf{Q}_{a,b}}^{\Delta/D,T}(\mathbf{r})$ between $\phi_{\mathbf{Q}_{a,b}}^{\Delta/D}(\mathbf{r})$ and $\phi_{\mathbf{Q}_{a,b}}^{T}(\mathbf{r})$ reflects the relative shift between superconductivity modulations and atomic topography. Figure S9 shows a typical example of the relative phase between $\phi_{\mathbf{Q}_{a,b}}^{D_1}(\mathbf{r})$ and $\phi_{\mathbf{Q}_{a,b}}^{T}(\mathbf{r})$. The statistical distributions of relative phases $\delta\phi_{\mathbf{Q}_{a,b}}^{D_1,T}(\mathbf{r})$ in both $a$ and $b$ directions peak around $\pm\pi$, showing the antiphase relation between $D_1$ and the atomic topography height of the topmost Te/Se lattice plane.

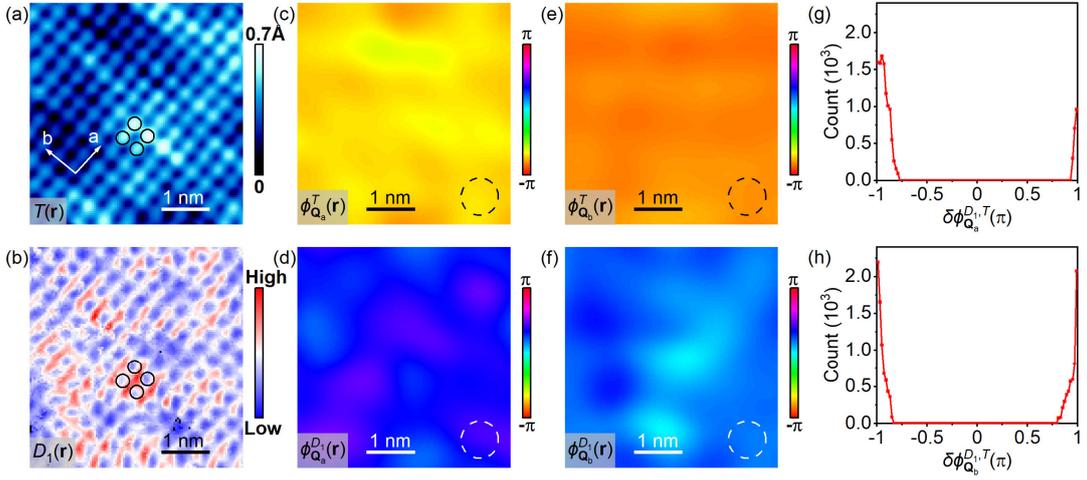

**Fig. S9.** Relative phase extraction. (a,b) The topographic image $T(\mathbf{r})$ and the coherence peak sharpness map $D_1(\mathbf{r})$ of Region 1 shown in Fig. S16 (4.4×4.6 nm$^2$, $T = 4.3$ K, $V_s = 40$ mV, $I_s = 500$ pA, $V_{\mathrm{mod}} = 0.8$ mV). The topmost Te/Se sites within one unit cell are marked by black circles in (a) and (b), indicating the antiphase relation between $D_1(\mathbf{r})$ and $T(\mathbf{r})$. (c,d) Phase maps of $T(\mathbf{r})$ and $D_1(\mathbf{r})$ of modulation at $\mathbf{Q}_a$. (e,f) Phase maps of $T(\mathbf{r})$ and $D_1(\mathbf{r})$ of modulation at $\mathbf{Q}_b$. The window sizes in $\mathbf{r}$-space are denoted by dashed circles in (c)-(f). (g,h) Distributions of the relative phase $\delta\phi_{\mathbf{Q}_i}^{D_1,T}(\mathbf{r})$ between $\phi_{\mathbf{Q}_i}^{D_1}(\mathbf{r})$ and $\phi_{\mathbf{Q}_i}^{T}(\mathbf{r})$, further suggesting the overall antiphase relation ($\delta\phi \approx \pm\pi$) between $T(\mathbf{r})$ and $D_1(\mathbf{r})$.

The stability of the phase analysis is checked by calculating the relative phase distributions using different $\sigma_q$ as shown in Fig. S10. The relative phase distribution is almost identical for different $\sigma_q$, demonstrating the reliability of our phase analysis.

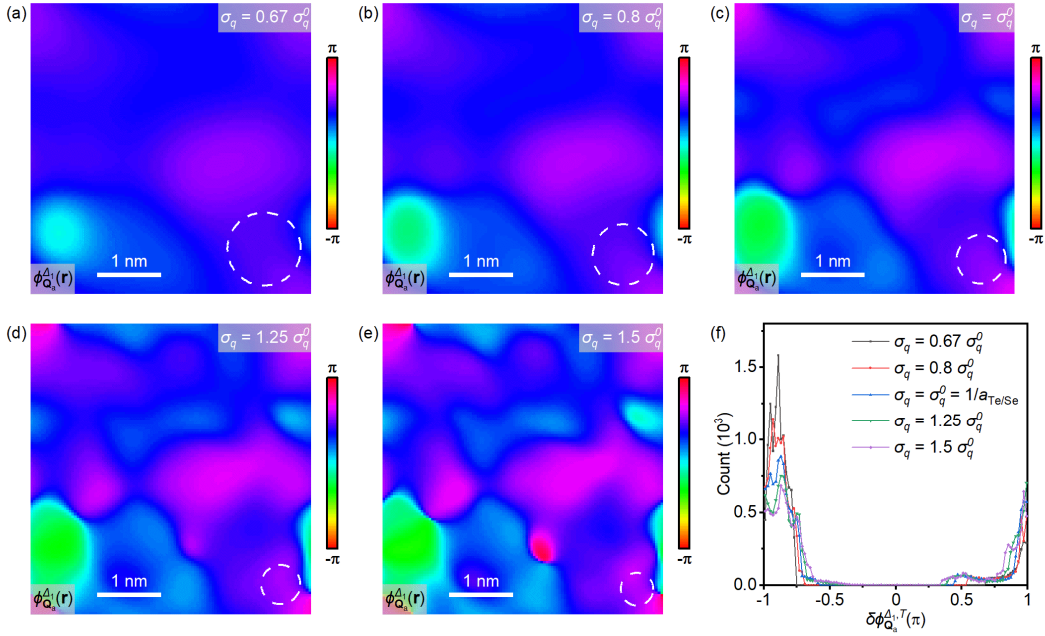

**Fig. S10.** Phase maps of $Q_a$ modulation of $\Delta_1$ with different window sizes. (a)-(e) Spatial phase map $\phi_{Q_a}^{\Delta_1}(\mathbf{r})$ of $\Delta_1$ modulation at $\mathbf{Q}_a$ in the same area shown in Fig. 3 calculated with increasing $\mathbf{q}$-space cut-off lengths $\sigma_q$ (4.4×4.6 nm², $T$ = 4.3 K, $V_s$ = 40 mV, $I_s$ = 500 pA, $V_{mod}$ = 0.8 mV). The window sizes in $\mathbf{r}$-space are denoted by dashed circles. (f) The distributions of the relative phase between superconducting gap size $\Delta_1$ and atomic topography calculated with different window sizes. All distributions are almost identical, which shows that the window sizes do not affect the obtained phase relationship. The $\mathbf{q}$-space cut-off length we used in our work is $\sigma_q^0 = 1/a_{Te/Se}$, which has relatively high resolutions both in $\mathbf{r}$-space and $\mathbf{q}$-space.

## Fourier filtering process

The Fourier filtering was performed by multiplying Gaussian windows centered at Fourier peaks $\mathbf{Q}_i$ in the Fourier map. Then take the inverse Fourier transform of the masked Fourier map to get the filtered map. The process is realized as below:

$$C^{\mathbf{Q}}(\mathbf{r}) = F^{-1}\left[ F[C(\mathbf{r})] \cdot \sum_i \frac{1}{2\pi\sigma_q^2} e^{-\frac{(\mathbf{q}-\mathbf{Q}_i)^2}{2\sigma_q^2}} \right]$$

where $C(\mathbf{r})$ is the real space map. In practice, $\mathbf{Q}_i$ are equal to Bragg peaks of top Te/Se lattice and $\sigma_q$ is $1/a_{Te/Se}$.

## Robust superconducting gap modulation within the vortex

As shown in Fig. S11, the superconducting gap modulation persists within the vortex, showing that the modulation is robust under the magnetic field. Applying the magnetic field can suppress the coherence of the Cooper-pair condensate. However, the superconducting pairing is present even inside the vortex, and the gap size characterizes the pairing strength. Thus, the gap modulation could persist in the vortex. The superconducting modulations with unchanged periodicity in vortices have ever been observed in other superconductors, such as the kagome superconductor CsV₃Sb₅ [7].

Furthermore, in cuprate superconductor $Bi_2Sr_2CaCu_2O_{8+x}$, the superconducting gap modulations with the $8a_0$ ($a_0$ is the distance between neighboring Cu atoms) periodicity have been observed both in vortices [8] and without magnetic fields [9].

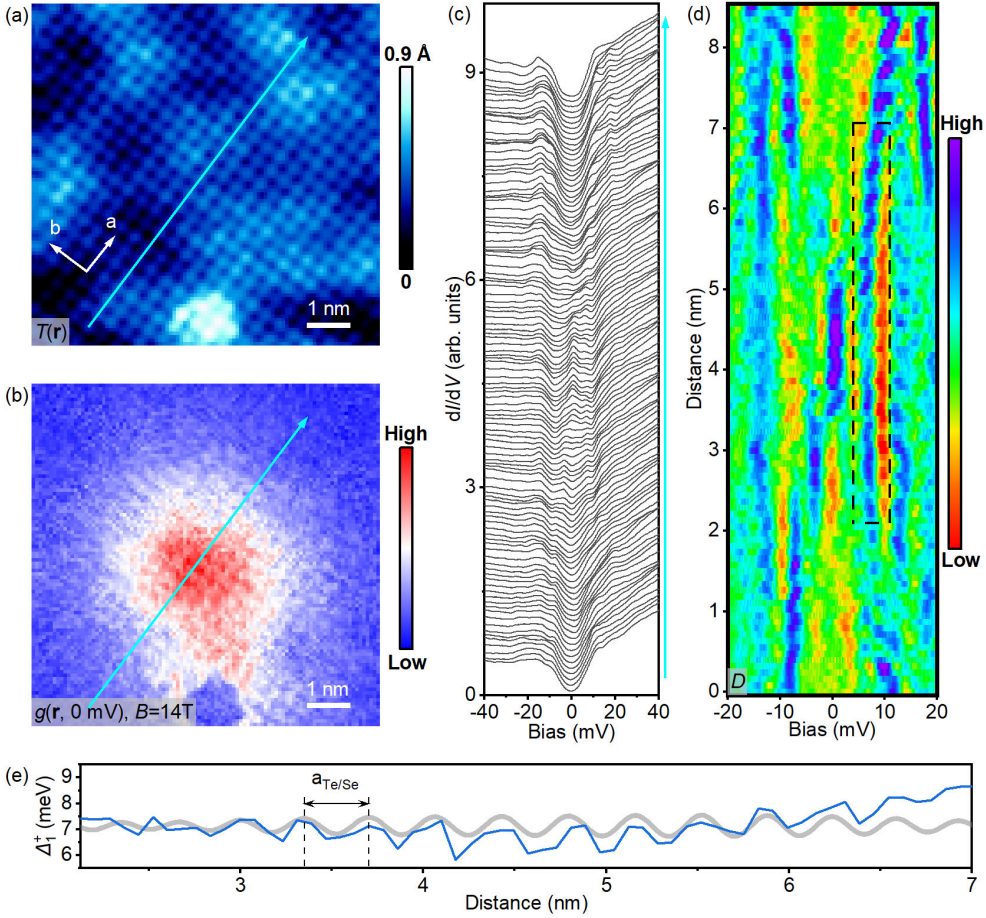

**Fig. S11.** Robust intra-unit-cell superconducting gap modulation across a vortex in a high magnetic field (Sample 1). (a) A topographic image of the 1-UC Fe(Te,Se)/STO (7.9×7.8 nm², $V_s$ = 40 mV, $I_s$ = 500 pA). (b) Zero bias conductance map $g(\mathbf{r}, 0\text{ mV}) \equiv dI/dV(\mathbf{r}, V = 0\text{ mV})$ measured in the same area of (a) under a $c$-axis magnetic field of $B$ = 14 T ($T$ = 4.3 K, $V_s$ = 40 mV, $V_{mod}$ = 0.8 mV). A single vortex is clearly revealed by higher zero bias conductance. The coherence length $\xi \approx$ 1.7 nm is estimated from the vortex core size. (c) Tunneling spectra $g \equiv dI/dV$ measured along the cyan arrow in (a) and (b) at $B$ = 14 T ($T$ = 4.3 K, $V_s$ = 40 mV, $I_s$ = 500 pA, $V_{mod}$ = 0.8 mV). The curves are vertically shifted for clarity. The cut line is chosen by the top Te/Se lattice direction across the vortex. Both the gap edges and the bound states in the vortex core are visible. (d) Color map of $D \equiv -d^2g/dV^2$ calculated from (c), which exhibits the spatially modulated superconducting gap size. (e) The extracted superconducting gap size (purple or blue colors) at positive bias ($\Delta_1^+$) within the dashed rectangle in (d). The gap sizes $\Delta_1^+$ exhibit spatial modulation along the top Te/Se lattice with a period of $a_{Te/Se}$. The gray curve is the extracted gap size modulation by applying the Fourier filter.

**Exclusion of possible influence from the setup effect**

We measured the superconducting gap modulations on 1-UC Fe(Te,Se)/STO with different setup conditions as shown in Fig. 3 ($V_s$ = 40 mV, $I_s$ = 500 pA), Fig. S8 ($V_s$ = 30 mV, $I_s$ = 8 nA) and Fig. S12 ($V_s$ = 70 mV, $I_s$ = 8 nA). The detected intra-unit-cell superconducting gap modulation results at different setup conditions are consistent with each other, which demonstrates the setup effect in gap maps does not influence the observed modulations.

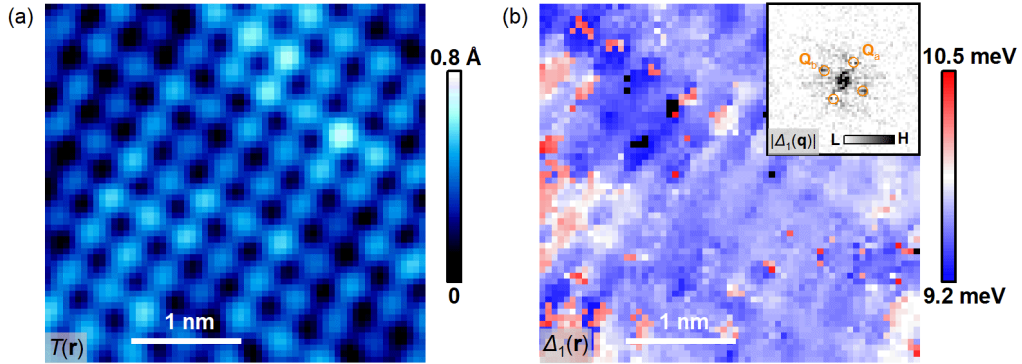

**Fig. S12.** Intra-unit-cell superconducting gap modulation measured at a higher setup voltage (Sample 3). (a) A topographic image of the 1-UC Fe(Te,Se)/STO showing the atomic structure of the topmost Te/Se lattice (3.5×3.3 nm², $V_s$ = 70 mV, $I_s$ = 8 nA). (b) Superconducting gap map of $\Delta_1$ extracted from $D$ curves in the same area as in (a), which shows the intra-unit-cell superconducting gap modulations ($T$ = 62 mK, $V_s$ = 70 mV, $I_s$ = 8 nA, $V_{mod}$ = 0.2 mV, 333 pixels/nm²). The superconducting gaps are extracted from the coherence peak structures at positive bias. Black pixels represent no well-defined peak structures at those positions. The inset of (b) is the magnitude of the Fourier transform of (b). The intra-unit-cell superconducting gap modulation wavevectors ($\mathbf{Q}_a$ and $\mathbf{Q}_b$) are denoted by orange circles, which are at the Bragg points of the topmost Te/Se lattice.

## Discussion on projection of Bloch wavefunctions

The Bloch wavefunctions of a Fermi surface with multi-orbital contributions may project non-uniformly onto different atomic sites, which may lead to modulations with the lattice periodicity. However, the projection usually can only lead to variations in the spectral intensities, i.e., different tunneling conductance at different atomic sites, but not the modulation of the superconducting energy gap. Thus, the most convincing evidence for the superconducting modulations lies in the observation of the superconducting gap size modulations, rather than those of the spectral intensities such as the coherence peak heights. For iron-based superconductors, the Fermi surface contains more contributions from the Fe $d$-orbital and less from the Te/Se $p$-orbital near the Fermi level. The tunneling conductance close to the Fermi energy will be higher on the Fe sites than on the Te/Se sites, but the superconducting gap size determined by the energy separation between the coherence peaks usually remains the same. The physics can also be seen from the perspective of the superconducting proximity effect that the $d$-electrons and $p$-electrons, with quantum mixing and in close proximity, should have the same superconducting gap. Thus, the modulation of the superconducting gap within the unit cell is an intriguing phenomenon that cannot be explained by the spectral weight modulations.

## Detailed discussion on superconducting gap modulations

Despite the intriguing modulation of the superconductivity detected between the top and bottom Te/Se locations in the unit cell, we observe no significant differences in the superconducting properties at the two Fe atoms in the unit cell. These findings provide microscopic evidence for the importance of the Te/Se atoms in the superconducting pairing interactions. This is supported by the theoretical proposal that the *p*-orbitals of the pnictogen and chalcogen can play a significant part in the pairing interactions mediated by electronic spin as well as charge and orbital fluctuations, in addition to the Fe *d*-orbital electrons in iron-based superconductors [10,11]. The strong *p-d* hybridization can enhance the effective antiferromagnetic coupling between the *d*-electrons of the next nearest neighboring Fe atoms [12–16]. The imbalance between the top and bottom Te/Se atoms, induced by the STO substrate, may lead to different pairing strengths and thus the observed PDM states. As shown schematically in Fig. S13, the next nearest neighbor pairing mediated by the top (bottom) Te/Se is represented by pairing bonds of different strength marked by yellow (orange), forming a staggered Fe-cornered checkboard lattice. Each Fe atom is connected with two strong pairing bonds and two weak pairing bonds, which prevents the Fe site from being at the modulation extremes of superconductivity. In contrast, the Te/Se atoms are centered at either two strong pairing bonds or two weak pairing bonds, leading to the superconductivity difference between the top and bottom Te/Se sites. Our observation of superconducting gap size modulation is overall consistent with this theoretical scenario.

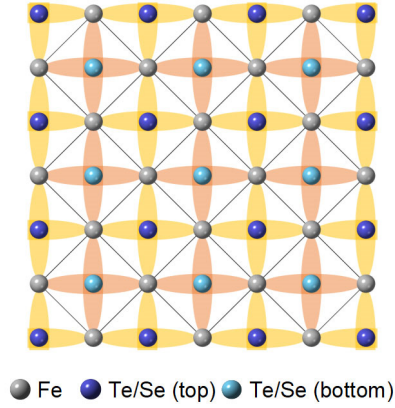

**Fig. S13.** Possible theoretical scenario for the PDM state. Schematic for the Te/Se assistant Cooper pairing between next nearest neighbor Fe atoms. Due to the STO substrate induced glide-mirror symmetry breaking, the top and bottom Te/Se are different, leading to different antiferromagnetic coupling strengths and pairing strengths (marked by yellow and orange crosses).

**Superconductivity modulation at Q = G on sublattices**

We would like to clarify the important physics of superconductivity on the crystal lattice with a sublattice structure. The meaning of uniform or Q = 0 superconductivity needs to be handled with some care, since all momentum defined on the lattice is modulo the lattice momentum G. So, Q = G is equivalent to Q = 0, due to the periodicity of the reciprocal lattice. Consider an explicit example of the simplest 1D lattice with a lattice constant *a*, then Q = 0, $2\pi/a$, $4\pi/a$ …, are all equivalent zero-momentum wave vectors, and we will keep only Q = 0 and Q = G = $2\pi/a$ for our discussion. In standard solid state theory, we fold everything to the first Brillouin zone, so that Q = G folds to Q = 0, and we need to just consider Q = 0. This works straightforwardly in an elemental solid with one atom per unit cell. For example, consider a density wave at Q: $\rho(r) = \rho_0 \cos(Q \cdot r + \varphi)$, where r labels the spatial location of an atomic site, which is simply the unit cell in this case. Clearly, $Q = 0$ and $Q = G$ return the same constant $\rho(r) = \rho_0 \cos(\varphi)$, since $Q \cdot r = 2n\pi$.

Next, we consider a solid with more than one atom per unit cell, located at $r_i$ with $i = 1, 2, \ldots, n$, labeling the n sublattices. r and $a$ continue to denote the unit cell location and the lattice spacing. For a superconducting state with the Cooper-pair center of mass momentum Q (corresponding to pairing between k and $-k+Q$), the order parameter expectation $\Delta(r_i) = \Delta_0 \cos(Q \cdot r_i + \varphi)$. Now at $Q = 0$, $\Delta(r_i)$ is still completely uniform in space. But for $Q = G$, the pairing order parameter $\Delta(r_i) = \Delta_0 \cos(G \cdot r_i + \varphi)$, which is no longer uniform in the unit cell, but shows spatial variations among the n sublattice sites, i.e., a nontrivial form factor for $Q = 0$ superconducting state. This describes the PDM states directly observed by STM in real space.

Then, the $Q = G$ superconductivity can also be understood as a $Q = 0$ superconductivity with multiple orbitals/bands. When G is folded into the first Brillouin zone (BZ), the sublattice index $i$ transforms to an effective orbital/band index $\alpha = 1, \ldots, n$, and as a result, the site or sublattice dependence of the pairing order parameter translates into orbital dependence, i.e., $\Delta(r_\alpha) = \Delta_0 \cos(Q \cdot r_\alpha + \varphi) = \Delta_0^\alpha$, which provides an equivalent description of the PDM.

Specifically for monolayer high-$T_c$ Fe(Te,Se) on STO substrate, consider the BZ containing one Fe (and one Te/Se) per unit cell plotted in Fig. S14 (blue square) together with the Fermi surface of hybridized Fe-$d$ and Te/Se-$p$ orbitals [10,11]. The spatial modulations of the superconducting gap have a periodicity of $\sqrt{2}a_{Fe} \times \sqrt{2}a_{Fe}$, which corresponds to $Q = (\pi,\pi)$ and equivalent points of the BZ. This indicates there is a component of the superconductivity connecting (green arrow in Fig. S14) different Fermi surface pockets that overlap at the four points (hotspots) by the $(\pi,\pi)$ translation. Now the true unit cell of the atomic structure for Fe-based superconductors contains two Fe atoms (and two Te/Se atoms) due to the staggered distribution of the Te/Se atoms above and below the Fe-plane. The BZ corresponding to the two-Fe unit cell is also shown in Fig. S14 (red square). As indicated in Fig. S14, the $Q = (\pi,\pi)$ wave vector becomes $Q = G$, i.e. the reciprocal lattice vector of the reduced two-Fe BZ, which can be folded to the $Q = 0$ $\Gamma$ point, together with the folding of the Fermi surface pockets (dashed lines). This leads to twice as many orbitals/bands or Fermi surfaces inside the two-Fe zone and the intersecting points of the original and the folded Fermi pockets are precisely at the "hotspots" connecting the original pockets by the $Q = (\pi,\pi)$ wave vector. Thus, in sum, the observed PDM amounts to a $Q = (\pi,\pi)$ modulation that connects different original Fermi pockets or a $Q = 0$ modulation in the folded zone between the original and the folded Fermi pockets as shown in Fig. S14, both reflect a real space modulation with the periodicity of $\sqrt{2}a_{Fe} \times \sqrt{2}a_{Fe}$.

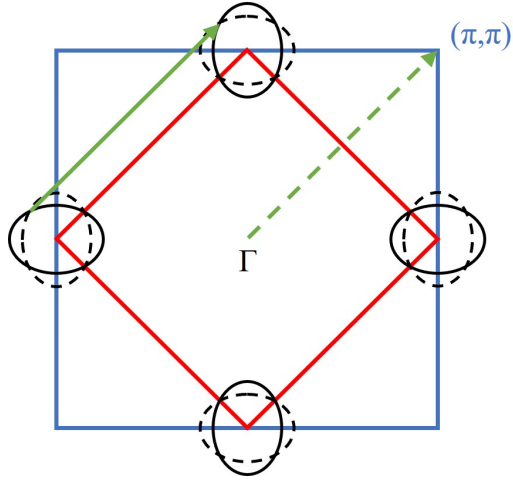

**Fig. S14.** Schematic of the Fermi surface of 1-UC Fe(Te,Se)/STO. Blue and red squares represent one-Fe BZ and two-Fe (folded) BZ, respectively. The original and folded Fermi surfaces are plotted by solid and dashed black lines, respectively. The solid green arrow is a $(\pi,\pi)$ wavevector connecting the hotspots on two different Fermi pockets in the one-Fe BZ, which is equal to the reciprocal lattice vector G (dashed green arrow) of the two-Fe BZ.

For a normal BCS superconductor, the superconducting coherence length is enormously longer than the lattice spacing, and it would be hard to observe superconducting gap modulations on the atomic scale. The PDM state in 1-UC Fe(Te,Se)/STO is unconventional and unprecedented, and requires at the minimum a short superconducting coherence length such that the sublattice structures become significant for superconductivity. This uncovers brand new physics in unconventional high-$T_c$ superconductors and offers new insights and directions to understanding the symmetry breaking effects, the local electronic structure, and the microscopic origin of the superconductivity.

**Phase relation between $\Delta(\mathbf{r})$ and $D(\mathbf{r})$ maps**

As shown in Figs. 3-4, an in-phase relation is observed between superconducting gap $\Delta_1$ and coherence peak sharpness $D_1$, while $\Delta_2$ and $D_2$ show the anti-phase relation. The relationship between the superconducting gap size and the coherence peak sharpness is an open question in unconventional superconductors. Generally speaking, it depends on the electronic structure such as the number of active bands, the correlation strength, the role of charge carriers, the pairing gap symmetry and momentum space anisotropy, etc. The only concrete understanding of the relationship is in the cuprates, which is a doped Mott insulator with a single band crossing the Fermi level. As the doping increases, the coherence peak becomes higher and sharper, while the superconducting gap decreases, suggesting an anti-phase relationship [17–20]. In the 1-UC Fe(Te,Se)/STO, the coexistence of two superconducting gaps and the influence of the STO substrate make this issue much more complex. There is still no concrete understanding of the phase relation between $\Delta(\mathbf{r})$ and $D(\mathbf{r})$ of 1-UC Fe(Te,Se)/STO currently. It is speculated that these two different relationships may arise from the anisotropic gap structures of 1-UC Fe(Te,Se)/STO. For example, the two gap energies may correspond to two different bands with different density of states behaviors. It is also possible that the two gaps of the 1-UC Fe(Te,Se)/STO originate from two extrema of an anisotropic superconducting gap. The coherence peak height or sharpness is related to the density of states of the Bogoliubov quasiparticle band at the coherence peak energy, which is inversely proportional to the band curvature. In this case, at the small $\Delta_1$, a larger gap value may correspond to a smaller band curvature and hence a sharper coherence peak. On the other hand, at the large $\Delta_2$, the larger gap value may correspond to larger band curvature and softer coherence peak. However, this explanation is based on the hypothesis on the actively debated origins of the two superconducting gaps in 1-UC Fe(Te,Se)/STO. Our work will inspire more researches to study this phenomenon, which may provide new insights into the Fe-based superconductors including 1-UC Fe(Te,Se)/STO.

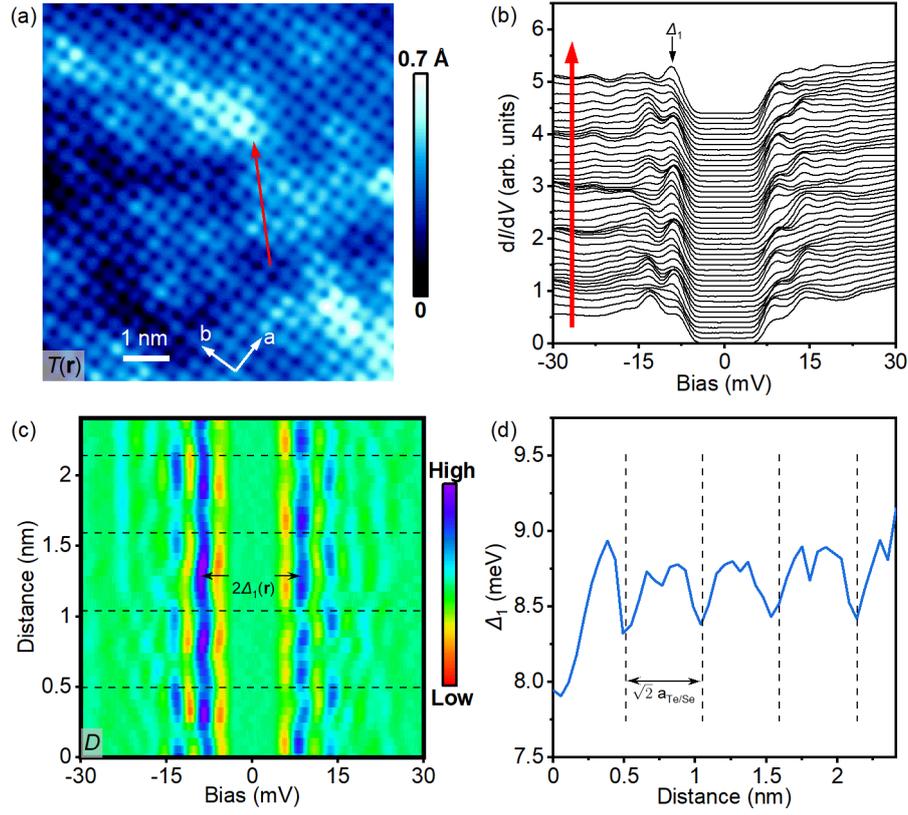

**Fig. S15.** Superconductivity modulation along Se/Te(top)-Se/Te(bottom) direction in 1-UC Fe(Te,Se) film (Sample 1). (a) A topographic image of the 1-UC Fe(Te,Se) film (7.6×8.3 nm², $V_s$ = 40 mV, $I_s$ = 500 pA). (b) Tunneling spectra measured along the red arrow in (a) ($T$ = 4.3 K, $V_s$ = 40 mV, $I_s$ = 500 pA, $V_{mod}$ = 0.8 mV). The curves are vertically shifted for clarity. The small black arrow indicates the small superconducting gap $\Delta_1$. (c) Color map of $D(V)$ = -d²$g$/d$V$² calculated from (b), which exhibits the spatially modulated superconducting gap size $\Delta_1$. (d) The extracted superconducting gap $\Delta_1$ along the distance in (c) showing spatial modulation with the period of $\sqrt{2}a_{Te/Se}$. The distances in (c) and (d) are defined relative to the beginning of the arrow in (a).

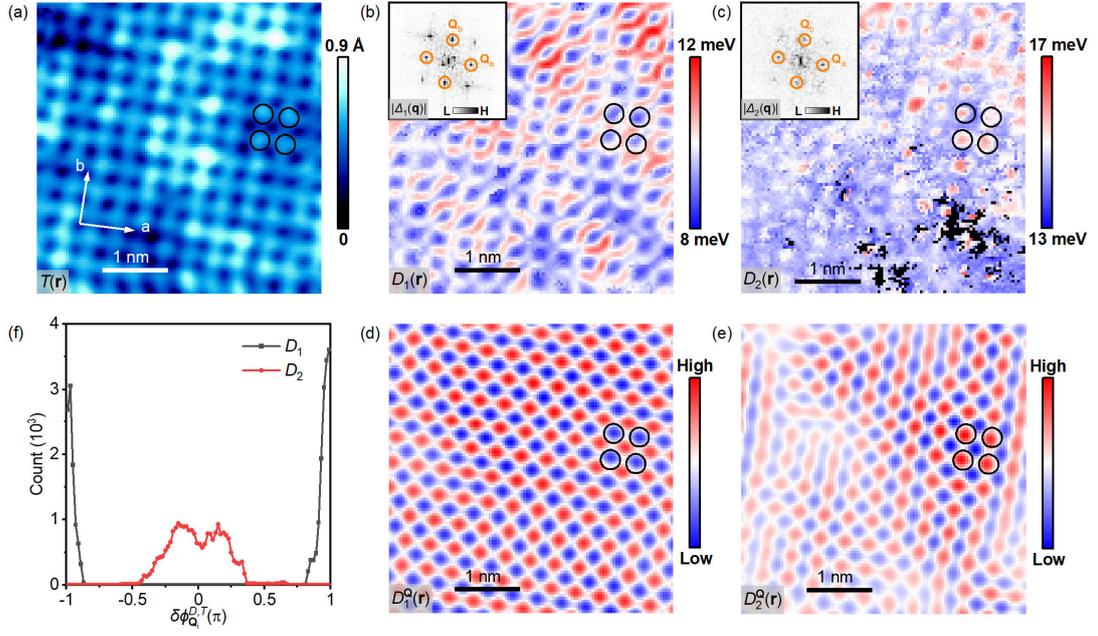

**Fig. S16.** Intra-unit-cell superconducting coherence peak sharpness modulation in the 1-UC FeSe/STO film. (a) Topographic image in Fig. 5(a). (4.3×4.4 nm$^2$, $V_s$ = 40 mV, $I_s$ = 500 pA) (b,c) Superconducting coherence peak sharpness maps of $D_1$ (b) and $D_2$ (c) measured in the same area as in (a), which show the intra-unit-cell $D$ modulation ($T$ = 4.3 K, $V_s$ = 40 mV, $I_s$ = 2 nA, $V_{mod}$ = 0.8 mV). The insets of (b) and (c) are the magnitude of the Fourier transform of (b) and (c), respectively. The intra-unit-cell coherence peak sharpness modulation wavevectors ($\mathbf{Q}_a$ and $\mathbf{Q}_b$) are denoted by orange circles, which are at the Bragg points of the topmost Se lattice. (d,e) The Fourier filtered $D$ maps of $D_1$ (d) and $D_2$ (e). The Fourier filter process only keeps the Fourier peaks around $\mathbf{Q}_a$ and $\mathbf{Q}_b$. The topmost Se sites within one unit cell are marked by black circles in (a)-(e). (f) The distributions of the relative phase between superconducting coherence peak sharpness ($D_1$ in black or $D_2$ in red) and atomic topography. The relative phase distributions peak near $\pm\pi$ (0) for $D_1$ ($D_2$), approximately showing antiphase (in-phase) relation between the $D_1$ ($D_2$) and the atomic topography.

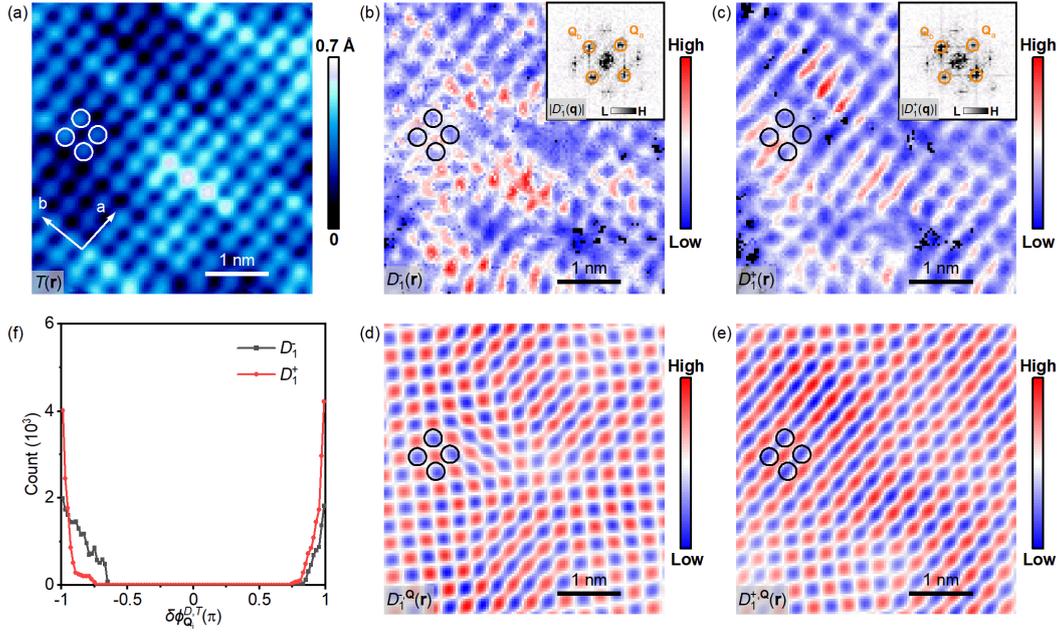

**Fig. S17.** In-phase superconducting coherence peak sharpness modulations at negative and positive biases. (a) The topographic image shown in Fig. 3(a) (4.4×4.6 nm$^2$, $V_s$ = 40 mV, $I_s$ = 500 pA). (b,c) Superconducting coherence peak sharpness maps of $D_1$ extracted at negative (b) and positive (c) biases in the same area as in (a), which show the intra-unit-cell superconducting coherence peak sharpness modulations ($T$ = 4.3 K, $V_s$ = 40 mV, $I_s$ = 500 pA, $V_{mod}$ = 0.8 mV). The insets of (b) and (c) are the magnitude of the Fourier transform of (b) and (c), respectively. (d,e) The Fourier filtered coherence peak sharpness maps of $D_1^-$ (d) and $D_1^+$ (e). The topmost Te/Se sites within one unit cell are marked by circles in (a)-(e), clearly revealing the antiphase relation between $D_1^{-,+}(\mathbf{r})$ and $T(\mathbf{r})$. (f) The distributions of the relative phase between superconducting coherence peak sharpness ($D_1^-$ in black or $D_1^+$ in red) and atomic topography. The relative phase distributions are concentrated near ±π, approximately showing that both $D_1^-$ and $D_1^+$ show antiphase relation with topography and the modulations of $D_1^-$ and $D_1^+$ are in-phase with each other.

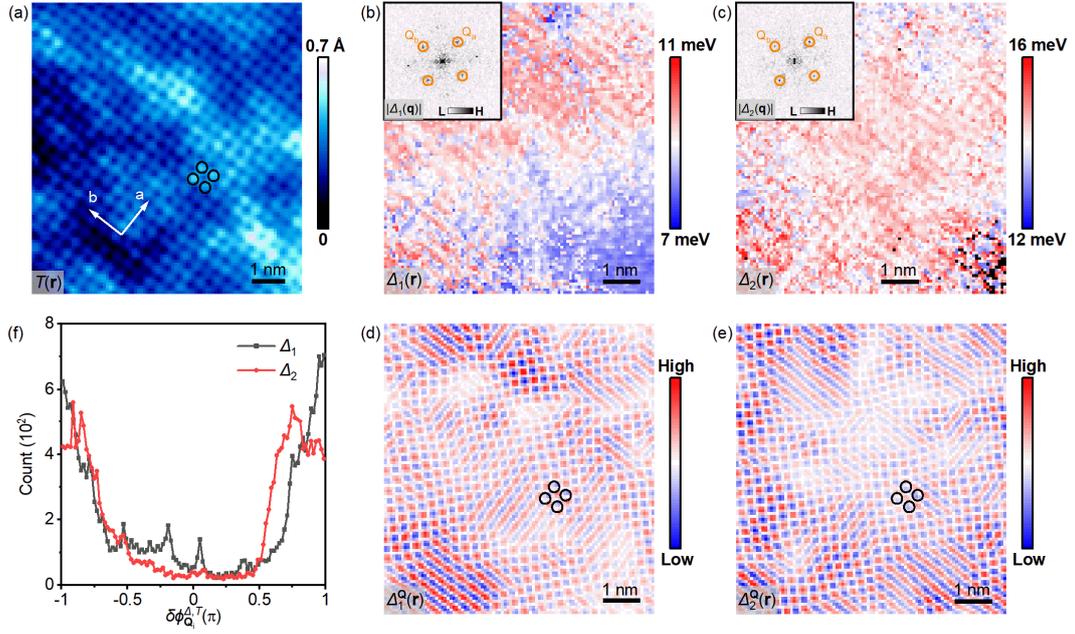

**Fig. S18.** Intra-unit-cell superconducting gap modulation in Sample 1. (a) A topographic image of Sample 1 (7.6×8.3 nm², $V_s$ = 40 mV, $I_s$ = 500 pA). (b,c) Superconducting gap maps of $\Delta_1$ (b) and $\Delta_2$ (c) measured in the same area as in (a), which show the intra-unit-cell superconducting gap modulations ($T$ = 4.3 K, $V_s$ = 40 mV, $I_s$ = 500 pA, $V_{mod}$ = 0.8 mV, 156 pixels/nm²). The insets of (b) and (c) are the magnitude of the Fourier transform of (b) and (c), respectively. The intra-unit-cell superconducting gap modulation wavevectors ($\mathbf{Q}_a$ and $\mathbf{Q}_b$) are denoted by orange circles, which are at the Bragg points of the topmost Te/Se lattice. (d,e) The Fourier filtered gap maps of $\Delta_1$ (d) and $\Delta_2$ (e). The topmost Te/Se sites within one unit cell are marked by black circles in (a), (d) and (e), revealing the antiphase relation between $\Delta_{1,2}(\mathbf{r})$ and $T(\mathbf{r})$. (f) The distributions of the relative phase between superconducting gap size ($\Delta_1$ in black or $\Delta_2$ in red) and atomic topography. The relative phase distributions are closer to ±π than to 0, approximately showing the antiphase relation between the gap size and the atomic topography.

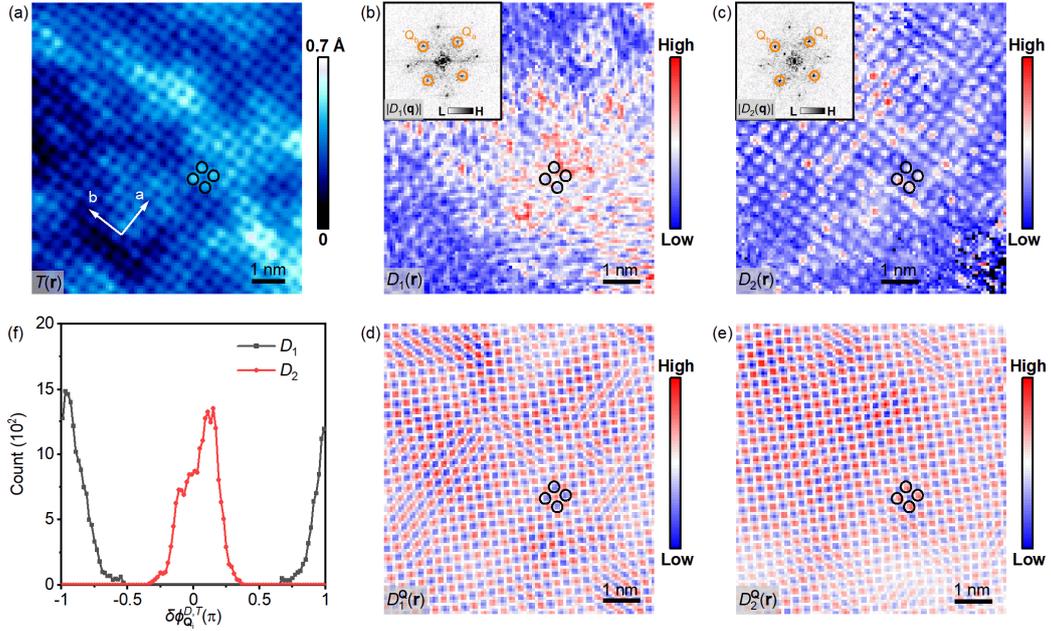

**Fig. S19.** Intra-unit-cell superconducting coherence peak sharpness modulation in Sample 1. (a) A topographic image of Sample 1 (7.6×8.3 nm², $V_s$ = 40 mV, $I_s$ = 500 pA). (b,c) $D_1$ (b) and $D_2$ (c) maps measured in the same area as in (a), which show the intra-unit-cell modulations ($T$ = 4.3 K, $V_s$ = 40 mV, $I_s$ = 500 pA, $V_{mod}$ = 0.8 mV, 156 pixels/nm²). The insets of (b) and (c) are the magnitude of the Fourier transform of (b) and (c), respectively. The intra-unit-cell coherence peak sharpness modulation wavevectors ($\mathbf{Q}_a$ and $\mathbf{Q}_b$) are denoted by orange circles, which are at the Bragg points of the topmost Te/Se lattice. (d,e) The Fourier filtered gap maps of $D_1$ (d) and $D_2$ (e). The topmost Te/Se sites within one unit cell are marked by black circles in (a)-(e), clearly revealing the antiphase relation between $D_1(\mathbf{r})$ and $T(\mathbf{r})$ and in-phase relation between $D_2(\mathbf{r})$ and $T(\mathbf{r})$. (f) The distributions of the relative phase between superconducting coherence peak sharpness ($D_1$ in black or $D_2$ in red) and atomic topography. The relative phase distributions are concentrated near $\pm\pi$ (0) for $D_1$ ($D_2$), showing antiphase (in-phase) relation between $D_1$ ($D_2$) and the atomic topography.

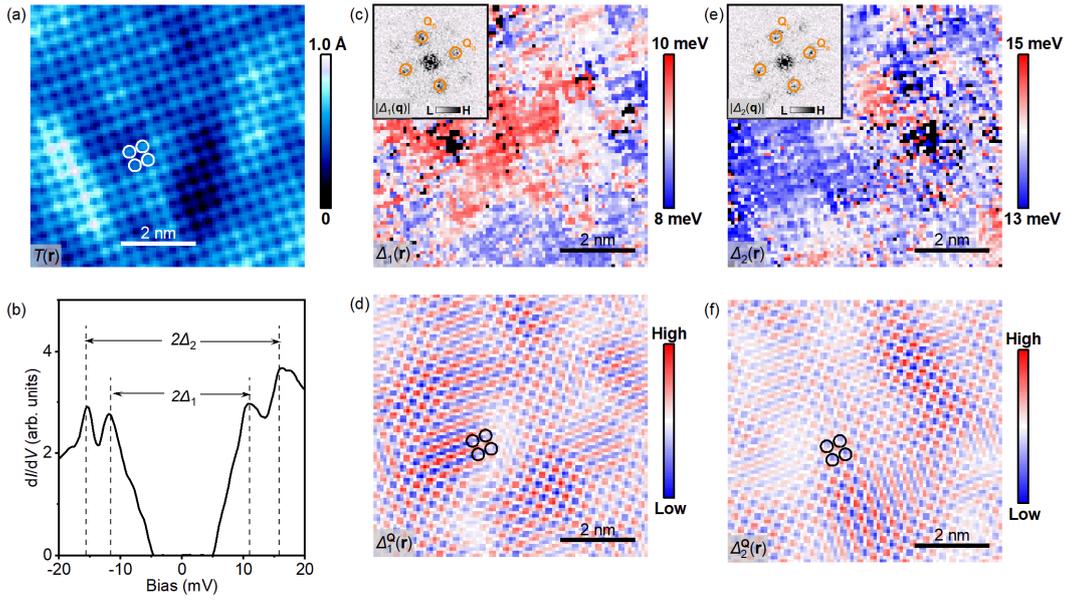

**Fig. S20.** Intra-unit-cell superconducting gap modulation in the 1-UC Fe(Te,Se) film (Sample 2) observed at 40 mK. (a) A topographic image of Sample 2 showing the atomic structures of the topmost Te/Se lattice ($7.3 \times 7.0$ nm$^2$, $V_s$ = 40 mV, $I_s$ = 4 nA). (b) A tunneling spectrum of Sample 2 with two superconducting gap $\Delta_1$ and $\Delta_2$ measured at 40 mK ($T$ = 40 mK, $V_s$ = 40 mV, $I_s$ = 4 nA, $V_{mod}$ = 0.4 mV). (c) Superconducting gap map of $\Delta_1$ measured in the same area as in (a), which shows the intra-unit-cell superconducting gap modulations ($T$ = 40 mK, $V_s$ = 40 mV, $I_s$ = 4 nA, $V_{mod}$ = 0.4 mV, 129 pixels/nm$^2$). The inset of (c) is the magnitude of the Fourier transform of (c). The intra-unit-cell superconducting gap ($\Delta_1$) modulation wavevectors ($\mathbf{Q}_a$ and $\mathbf{Q}_b$) are denoted by orange circles, which are at the Bragg points of the topmost Te/Se lattice. (d) The Fourier filtered gap map of $\Delta_1$. (e) Superconducting gap map of $\Delta_2$ measured in the same area as in (a) ($T$ = 40 mK, $V_s$ = 40 mV, $I_s$ = 4 nA, $V_{mod}$ = 0.4 mV, 129 pixels/nm$^2$). The inset of (e) is the magnitude of the Fourier transform of (e). The intra-unit-cell superconducting gap ($\Delta_2$) modulation wavevectors ($\mathbf{Q}_a$ and $\mathbf{Q}_b$) are denoted by orange circles, which are at the Bragg points of the topmost Te/Se lattice. (f) The Fourier filtered gap map of $\Delta_2$.

**References**

[1] F. Li et al., Interface-enhanced high-temperature superconductivity in single-unit-cell $FeTe_{1-x}Se_x$ films on $SrTiO_3$, Phys. Rev. B **91**, 220503 (2015).

[2] X.-L. Peng et al., Observation of topological transition in high-$T_c$ superconducting monolayer $FeTe_{1-x}Se_x$ films on $SrTiO_3(001)$, Phys. Rev. B **100**, 155134 (2019).

[3] J.-X. Yin et al., Observation of a robust zero-energy bound state in iron-based superconductor Fe(Te,Se), Nat. Phys. **11**, 543 (2015).

[4] J. R. Schrieffer, Theory of Electron Tunneling, Rev. Mod. Phys. **36**, 200 (1964).

[5] R. C. Dynes, V. Narayanamurti, and J. P. Garno, Direct Measurement of Quasiparticle-Lifetime Broadening in a Strong-Coupled Superconductor, Phys. Rev. Lett. **41**, 1509 (1978).

[6] Z. Du, X. Yang, H. Lin, D. Fang, G. Du, J. Xing, H. Yang, X. Zhu, and H.-H. Wen, Scrutinizing the double superconducting gaps and strong coupling pairing in $(Li_{1-x}Fe_x)OHFeSe$, Nat. Commun. **7**, 10565 (2016).

[7] H. Chen et al., Roton pair density wave in a strong-coupling kagome superconductor, Nature **599**, 222 (2021).

[8] S. D. Edkins et al., Magnetic field–induced pair density wave state in the cuprate vortex halo, Science **364**, 976 (2019).

[9] Z. Du, H. Li, S. H. Joo, E. P. Donoway, J. Lee, J. C. S. Davis, G. Gu, P. D. Johnson, and K. Fujita, Imaging the energy gap modulations of the cuprate pair-density-wave state, Nature **580**, 65 (2020).

[10] S. Zhou, G. Kotliar, and Z. Wang, Extended Hubbard model of superconductivity driven by charge fluctuations in iron pnictides, Phys. Rev. B **84**, 140505 (2011).

[11] Y. Yanagi, Y. Yamakawa, and Y. Ōno, Two types of $s$-wave pairing due to magnetic and orbital fluctuations in the two-dimensional 16-band $d$-$p$ model for iron-based superconductors, Phys. Rev. B **81**, 054518 (2010).

[12] I. I. Mazin, D. J. Singh, M. D. Johannes, and M. H. Du, Unconventional Superconductivity with a Sign Reversal in the Order Parameter of $LaFeAsO_{1-x}F_x$, Phys. Rev. Lett. **101**, 057003 (2008).

[13] K. Kuroki, S. Onari, R. Arita, H. Usui, Y. Tanaka, H. Kontani, and H. Aoki, Unconventional Pairing Originating from the Disconnected Fermi Surfaces of Superconducting $LaFeAsO_{1-x}F_x$, Phys. Rev. Lett. **101**, 087004 (2008).

[14] A. V. Chubukov, D. V. Efremov, and I. Eremin, Magnetism, superconductivity, and pairing symmetry in iron-based superconductors, Phys. Rev. B **78**, 134512 (2008).

[15] P. J. Hirschfeld, M. M. Korshunov, and I. I. Mazin, Gap symmetry and structure of Fe-based superconductors, Rep. Prog. Phys. **74**, 124508 (2011).

[16] R. M. Fernandes, A. I. Coldea, H. Ding, I. R. Fisher, P. J. Hirschfeld, and G. Kotliar, Iron pnictides and chalcogenides: a new paradigm for superconductivity, Nature **601**, 35 (2022).

[17] S. H. Pan et al., Microscopic electronic inhomogeneity in the high-$T_c$ superconductor $Bi_2Sr_2CaCu_2O_{8+x}$, Nature **413**, 282 (2001).

[18] Z. Wang, J. R. Engelbrecht, S. Wang, H. Ding, and S. H. Pan, Inhomogeneous $d$-wave superconducting state of a doped Mott insulator, Phys. Rev. B **65**, 064509 (2002).

[19] Y. Kohsaka et al., How Cooper pairs vanish approaching the Mott insulator in $Bi_2Sr_2CaCu_2O_{8+\delta}$, Nature **454**, 1072 (2008).

[20] S. Ye, C. Zou, H. Yan, H. Ji, M. Xu, Z. Dong, Y. Chen, X. Zhou, and Y. Wang, The emergence

of global phase coherence from local pairing in underdoped cuprates, Nat. Phys. **19**, 1301 (2023).